\documentclass[twocolumn]{autart}    

\usepackage{graphicx}          
\usepackage{amsmath}
\usepackage{cite}
\usepackage{caption}
\usepackage{subcaption}

\usepackage[font=scriptsize]{caption}

\usepackage{txfonts}

\begin{document}

\begin{frontmatter}

\title{Attack Analysis for Distributed Control Systems:\\ An Internal Model Principle Approach} 

\author[RM]{Rohollah Moghadam}\ead{moghadamr@mst.edu} and    
\author[RM]{Hamidreza Modares}\ead{modaresh@mst.edu}               

\address[RM]{Department
of Electrical and Computer Engineering, Missouri University of Science and Technology, USA}  


\begin{abstract}                          
Although adverse effects of attacks have been acknowledged in many cyber-physical systems, there is no system-theoretic comprehension of how a compromised agent can leverage communication capabilities to maximize the damage in distributed multi-agent systems. A rigorous analysis of cyber-physical attacks enables us to increase the system awareness against attacks and design more resilient control protocols. To this end, we will take the role of the attacker to identify the worst effects of attacks on root nodes and non-root nodes in a distributed control system. More specifically, we show that a stealthy attack on root nodes can mislead the entire network to a wrong understanding of the situation and even destabilize the synchronization process. This will be called the internal model principle for the attacker and will intensify the urgency of designing novel control protocols to mitigate these types of attacks.
\end{abstract}

\end{frontmatter}

\section{Introduction}
A cyber-physical system (CPS) refers to a relatively new generation of systems that integrates  computation, networking, and physical processes. Based on the control objectives, CPSs can be classified into two categories. The first class, called the networked control system (NCS), is a large-scale, but single-agent distributed system, wherein sensors, actuators, and controllers are distributed across the system, and the control loops are closed through a real-time communication network \cite{Campbell2010,Gupta2010,sridhar2012cyber,HespanhaNCS2007}. The global objective in a NCS is to assure that the output of the system tracks a predefined trajectory or regulates to the origin. The second class, called the multi-agent system (MAS), which is the problem of interest in this paper, is composed of a set of dynamical agents that are interacting with each other to achieve a coordinated operation and behavior \cite{Olfati2007,Robbins2013,Bidram2014,Bullo2009,Cortes2004,Ogren2004}. Despite their numerous applications in a variety of disciplines, distributed MASs are vulnerable to CPS attacks. In contrast to other adversarial inputs, such as disturbances and noises, attacks are intentionally planned to maximize the harm the system or even destabilize it. To develop resilient control protocols for optimal risk mitigation, performance assurance, and survivability in uncertain and changing networked environments, a science for modeling of adversarial behaviors and threats is required.

Several results have been reported on the severe damages of CPS attacks on NCSs \cite{chabukswar2010simulation,east2009taxonomy,falliere2011w32,fovino2009experimental,assante2016confirmation,Zou2016,Long2005,shoukry2013non} and many approaches have been proposed for attack mitigation or detection in these systems \cite{Fawzi2014,Pajic2014,Pasqualetti2013,Shoukry2014Event,Mo2015,Muradore2015}. The effects of attacks on distributed MASs can be even more serious and difficult to detect compared to NCSs, as the source of the attack might be some compromised neighbors, not the agent itself. Despite its importance, there is no rigorous and mathematical framework to characterize conditions under which an attacker maximizes the damage to the network and, for example, destabilizes the synchronization or consensus in MASs. Attack detection and mitigation techniques for MASs have been  developed in the literature \cite{Pasqualetti2012TAC,Teixeira2010,Fagiolini2013,Mo2014,Amin2013TCST,Pang2012TCST,Weerakkody2017,Feng2017ACC,Khazraei2017ACC,Sundaram2011,Zeng2014Resilient}. Most of the mentioned mitigation approaches use the discrepancy among agents and their neighbors or the exact state of agents to detect and mitigate the effect of the attacker. However, we will show that  a stealthy attack on a root node can cause an emergent misbehavior in the network  with no discrepancy between agents' states, and thus, existing mitigation approaches do not work. Moreover, this discrepancy could be as a result of a legitimate change in the state of an agent and blindly rejecting its information can harm the network connectivity and, consequently, convergence of the network. Therefore, rigorous analysis of attacks on MASs is required to identify the worst effects of attacks and  highlight the urgency of designing novel resilient control protocols.

In this paper, we rigorously analyze the effects of attacks on a distributed MAS. It is first shown mathematically that adversarial inputs launched on a single agent can snowball into a much larger and more catastrophic one. Conditions under which an attacker can destabilize the entire synchronization are found. This is called the internal model principle for the attacker. The internal model principle is well established for distributed output synchronization of MASs \cite{Wieland2011,Xiang20171,Lunze2011}. It says that \cite{FRANCIS1976457} the system needs to incorporate the dynamics of the reference or disturbance into its dynamic to achieve perfect tracking or disturbance rejection. The attacker, on the other hand, can identify a root node and incorporate a natural mode of the agents dynamics into the design of its attack signal to destabilize the network. This natural mode can be obtained by eavesdropping and monitoring the sensory data without having knowledge of the agents dynamics. Conditions under which the local neighborhood tracking error becomes zero despite attack is also found. This shows that existing disturbance attenuation techniques \cite{basseville1993detection,Jiao2016361} that attempt to minimize the effect of disturbance on the local neighborhood tracking error do not work in the presence of stealthy attacks in which the attacker has the knowledge of the agent dynamics and the network topology. 

The rest of the paper is organized as follows. Section 2 presents some preliminaries on graph theory and consensus of distributed MAS. In Section 3, the distributed consensus of MASs under attack is analyzed using frequency response and graph-theoretic approaches. Section 4 uses the results of Section 3 to provide a thorough discussion on the adverse effects of attacks on distributed MASs. Finally, a simulation example and the conclusion are presented in Sections 5 and 6, respectively.
\section{Preliminaries}
In this section, the preliminaries of the graph theory and consensus of the distributed multi-agent system (MAS) are provided. 
\subsection{Graph Theory} 
A directed graph $\mathcal{G}$ consists of a pair $\left( {\mathcal{V},\mathcal{E}} \right)$ in which $\mathcal{V}{\text{ =  }}\{ {v_1}, \cdots ,{v_N}\} $ is a set of nodes and  $\mathcal{E} \subseteq \mathcal{V} \times \mathcal{V}$ is a set of edges. The adjacency matrix is defined as 
$E = \left[ {{e_{ij}}} \right]$, with 
${e_{ij}} > 0$  if 
$({v_j},{v_i}) \in \mathcal{E}$, and  
${e_{ij}} = 0$  otherwise. The set of nodes ${v_i}$ with edges incoming to node 
${v_j}$  is called the neighbors of node $v_i$ , namely
${\mathcal{N}_i} = \{ {v_j}:({v_j},{v_i}) \in \mathcal{E}\} $ . The graph Laplacian matrix is defined as $\mathcal{L}=D-E$, where $D = diag({d_i})$ is the in-degree matrix, with ${d_i} = \sum\nolimits_{j \in {N_i}} {{e_{ij}}} $ as the weighted in-degree of node $v_i$. A (directed) tree is a connected digraph where every node, except the root node, has the in-degree of one. A graph is said to have a spanning tree if a subset of edges forms a directed tree.

Throughout the paper, ${\lambda}_A$ denotes the eigenvalue of matrix $A$, the Kronecker product of matrices $A$ and $B$ is $A \otimes B$ and $diag\left( {{A_1},\dots,{A_n}} \right)$   represents a block diagonal matrix with matrices ${A_i}$ , ${\rm{ }}i = 1,\dots,n$ as its diagonal entries. Finally, ${{\mathbf{1}}_N}$ is the ${N}$-vector of ones and $\operatorname{Im} (R)$ and $\ker (R)$ denote the range space and the null space of $R$, respectively.
\medskip

\noindent
\textbf{Assumption 1.}\textit{ The communication graph has a spanning tree.}
\subsection{Standard Distributed Consensus in MASs}
In this subsection, consensus for a leaderless MAS is reviewed. Consider $N$ agents with identical dynamics given by
\begin{equation}\label{eq:1}
{{\dot x}_i} = A{x_i} + Bu_i\,, 
\end{equation}
where $x_i\in{\mathbb{R}^n}$ and $u_i\in{\mathbb{R}^m}$ denote the system state and the control input, respectively. Matrices $A$ and $B$ are the drift and input dynamics, respectively. It is assumed that $\left( {A,B} \right)$ is stabilizable. 

Define the distributed local state variable control protocol for each node as \cite{lewis2013cooperative}
\begin{equation}\label{eq:2}
{u_i} = cK\sum\limits_{j \in N_i} {{e_{ij}}\left( {{x_j} - {x_i}} \right)}\,, 
\end{equation}
where \(c > 0\) denotes the scalar coupling gain, $K \in {\mathbb{R}^{m \times n}}$ is the feedback control gain matrix and \({e_{ij}}\) is the \((i,j)\)-th entry of the graph adjacency matrix.

Using \eqref{eq:1} and \eqref{eq:2}, the global dynamic of agents becomes
\begin{equation}\label{eq:3} 
\dot x = \left( {{I_N} \otimes A - c\mathcal{L} \otimes BK} \right)x\,, 
\end{equation}
\noindent
where $x = {\left[ {x_1^T, \ldots ,x_N^T} \right]^T} \in {\mathbb{R}^{Nn \times 1}}$ and $\mathcal{L}$ denotes the graph Laplacian matrix. The solution of \eqref{eq:3} is obtained by
\begin{equation}\label{eq:4} 
x(t) = {e^{\left( {{I_N} \otimes A - c\mathcal{L} \otimes BK} \right)t}}x(0)\,. 
\end{equation}

Let $K$ be designed such that the matrices \(A-c{\lambda _i}BK,\,i = 2, \ldots ,N\) be Hurwitz, where \({\lambda _i}{\rm{ , }}\,i = 2, \ldots ,N\) are the nonzero eigenvalues of the graph Laplacian matrix \(\mathcal{L}\). Then, under Assumption 1, the final consensus value is\cite{ li2014cooperative}
\begin{equation}\label{eq:FC_ledless}
{x}(t) \buildrel \Delta \over = \left( {{w^T} \otimes {e^{At}}} \right)\left[ {\begin{array}{*{20}{c}}{{x_1}(0)}\\ \vdots \\{{x_N}(0)}\end{array}} \right]\quad as\quad {\rm{ }}t \to \infty\,,
\end{equation}
where \(w = {\left[ {{p_1}, \ldots ,{p_N}} \right]^T} \in {\mathbb{R}^N}\) is the left eigenvector of \(\mathcal{L}\) associated with the zero eigenvalue and satisfies ${w^T}{{\mathbf{1}}_N}  = 1$.
\medskip

\noindent
\textbf{Assumption 2.}\textit{ The system dynamic $A$ in \eqref{eq:1} is marginally stable, with all eigenvalues on the imaginary axis.}
\medskip

\section{Distributed Consensus of MASs under Attack}
In this section, attacks on distributed MASs are modeled and their adverse effects on the standard distributed control protocol are analyzed. Then, the internal model principle for the attacker is investigated. Finally, attack analysis based on graph theoretic tools is provided. 
\subsection{Attack Modeling and Analysis for Distributed MAS}
In this subsection, attacks on distributed MASs are modeled and their adverse effects are analyzed.

The actuator attack on agent $i$ can be modeled by
\begin{equation}\label{eq:8}
u_i^c = u_i^{} + {\alpha _i}\,u_i^a\,,  	
\end{equation}
where ${u_i} \in {\mathbb{R}^m}$ is the nominal control input, \(u_i^a \in {\mathbb{R}^m}\) is the attack signal injected into the actuators of the agent $i$, \(u_i^c\) is the corrupted input applied to the MAS \eqref{eq:1} and 
\[ {\alpha _i} = \left\{ \begin{array}{ll}
         1 & \mbox{\quad Agent $i$ is under actuator attack}\\
         0 & \mbox{\quad Otherwise}\end{array} \right.\,. \]

The sensor attack is modeled as
\begin{equation}\label{eq:9} 
x_i^c = {x_i} + {\beta _i}\,x_i^a\,,  	
\end{equation}
where ${x_i}\in{\mathbb{R}^n}$ is the normal state, $x_i^a\in{\mathbb{R}^n}$ represents the attack signal injected into the sensors of the agent $i$, $x_i^c$ is the corrupted state and
\[ {\beta _i} = \left\{ \begin{array}{ll}
         1 & \mbox{\quad Agent $i$ is under sensor attack}\\
         0 & \mbox{\quad Otherwise}\end{array} \right.\,. \]
         
Using \eqref{eq:8} and \eqref{eq:9} in \eqref{eq:1} and \eqref{eq:2} yields 
\begin{equation}\label{eq:sysa} 
{\dot x_i} = A{x_i} + B{u_i} + B{f_i}\,,
\end{equation}
where \({f_i}\) represents the overall attack injected into the agent \(i\) and is given by
\begin{equation}\label{eq:atts} 
{f_i} = {\alpha _i}u_i^a + cK\left( {\sum\limits_{j \in {N_i}} {{e_{ij}}\left( {{\beta _j}x_j^a - {\beta _i}x_i^a} \right)} } \right)\,.
\end{equation}
\textbf{Remark 1.} Deception attacks on communication links can also be modeled the same as attacks on sensors given by \eqref{eq:9} \cite{Teixeira2010}. For this case, the disturbance can be injected by an attacker once the information of the agent is transmitted to its neighbors through the communication network. It should also be noted that attacks on sensors and actuators can be launched without physical tampering with the system. For example, the global positioning system (GPS) of an unmanned vehicle can be spoofed or the communication channel from the controller to the actuator can be jammed.
\medskip

\begin{thm}
Consider the MAS \eqref{eq:sysa} with the control protocol \eqref{eq:2}. Let agent $j$ be compromised by the attacker. Then, the intact agents that are reachable from this agent are disrupted from the desired consensus value.
\end{thm}

\begin{pf}
The proof is similar to our proof of Theorem 1 in \cite{Rohollah2017CDC} and is omitted.\hfill $\blacksquare$
\end{pf}

\subsection{Internal Model Principle for the Attacker}
In this subsection, we derive an internal model principle for the attacker and analyze its adverse effects on the consensus in distributed MASs. We show that in the leaderless distributed MAS, an attacker can inject a state-independent attack into sensors or actuators of a single root node to destabilize the entire consensus process. These effects are analyzed using the frequency response of the MAS. A more rigorous analysis is presented in the next section.

We now show that the attacker can design its attack signal, instead of blindly injecting a disruptive signal, to maximize the harm to the network.

Let the attack signal for agent $i$ be generated by
\begin{equation}\label{eq:12} 
\dot f_i = Rf_i,
\end{equation}
where $R \in {\mathbb{R}^{m \times m}}$. Define the set of the eigenvalues of the system dynamic $A$ defined in \eqref{eq:1} and the set of the eigenvalues of the attacker dynamic $R$ as
\begin{equation}\label{eq:ErEa}
\begin{gathered}
  {E_A} = \{ {\lambda _{{A_1}}}, \ldots ,{\lambda _{{A_n}}}\}  \hfill \\
  {E_R} = \{ {\lambda _{{R_1}}}, \ldots ,{\lambda _{{R_m}}}\}  \hfill \\ 
\end{gathered} 
\end{equation}

\noindent
\textbf{Lemma 1.}\cite{lewis2013cooperative},\cite{wu2007synchronization} Let Assumption 1 be satisfied. Define the set of root nodes as \(W \subset \mathcal{V}\) and the left eigenvector of the graph Laplacian matrix \(\mathcal{L}\) for \({\lambda _1} = 0\) as \(w = {[p_1,\dots,p_N ]^T}\). Then, ${p_i} > 0$ if $i \in W$ and ${p_i} = 0$ if $i \notin W$.\hfill $\blacksquare$ \medskip

The following theorem shows under what conditions the attacker can destabilize the network or cause a non-emergent, but stable behavior. These conditions are exploited more in the next section and are related to attacks on root nodes and non-root nodes.
\medskip

\begin{thm}(\textbf{Internal Model Principle for Attacker})

\noindent
Consider the MAS \eqref{eq:1} with the control protocol defined as
\begin{equation}\label{eq:39}
{u_i} = cK\sum\limits_{j \in {N_i}} {{e_{ij}}\left( {{x_j} - {x_i}} \right)}  + {f_i}\,,
\end{equation}
with \({f_i}\) defined in \eqref{eq:12}. Let $K$ be designed such that $A-\lambda_iBK$ $\forall i=2,\dots,N$ become Hurwitz, i.e., agents reach consensus when $f_i=0$ $\forall i=1,\dots,N$. Then, injecting an attack signal into the root nodes, 

\begin{enumerate}
\item destabilizes the MAS \eqref{eq:1}, if ${E_R} \cap {E_A} \ne \emptyset$ and $\sum\nolimits_{j = 1}^N {{p_{1j}}{f_j}}\ne 0$ with $E_R$ and $E_A$ defined in \eqref{eq:ErEa}.
\item cannot destabilize the MAS \eqref{eq:1}, but causes a non-emergent behavior, if  $\sum\nolimits_{j = 1}^N {{p_{1j}}{f_j}}= 0$  or ${E_R} \cap {E_A} =\emptyset$.
\end{enumerate}
\end{thm}
\begin{pf}
Define the transfer function of the MAS \eqref{eq:1} from \({x_i}(s)\) to \({u_i}(s)\) as
\begin{equation}\label{eq:40}
{G}(s) = \frac{{{x_i}(s)}}{{{u_i}(s)}} = {\left( {sI - A} \right)^{ - 1}}B\,.
\end{equation}

From \eqref{eq:39}, the global control signal vector can be written as 
\begin{equation}\label{eq:41}
u(s) = -(c\mathcal{L} \otimes K) x(s) + f(s)\,,
\end{equation}
with $u = {\left[ {{u_1}^T,\dots,{u_N}^T} \right]^T}$, $x = {\left[ {{x_1}^T,\dots,{x_N}^T} \right]^T}$ and $f = {\left[ {{f_1}^T,\dots,{f_N}^T} \right]^T}$. The overall state in terms of the transfer function \eqref{eq:40} becomes
\begin{equation}\label{eq:42}
\begin{gathered}
  x(s) = ({I_N} \otimes G(s))u(s) \hfill \\
   = ({I_N} \otimes G(s))\left[ { - (c\mathcal{L} \otimes K)x(s) + f(s)} \right] \hfill \\ 
\end{gathered}\,\,.
\end{equation}

Let $V$ be a nonsingular matrix such that $\mathcal{L}=V\Lambda V^{-1}$, with $\Lambda$ the Jordan canonical form of $\mathcal{L}$. Since the right and left eigenvectors of $\mathcal{L}$ corresponding to the zero eigenvalue are $\mathbf{1}_N$ and $w$, respectively, we define
\[V = \left[ {\begin{array}{*{20}{c}}
  {\mathbf{1}}&{{Y_1}} 
\end{array}} \right],{V^{ - 1}} = \left[ {\begin{array}{*{20}{c}}
  {{w^T}} \\ 
  {{Y_2}} 
\end{array}} \right]\,,\]
with $Y_1 \in \mathcal{R}^{N \times (N-1)}$ and $Y_2 \in \mathcal{R}^{(N-1) \times N}$ \cite{li2014cooperative}. Using $\mathcal{L}=V\Lambda V^{-1}$ and $V^{-1}V=I_N$, \eqref{eq:42} turns into
\begin{equation} \label{eq:42new}
 \begin{gathered} 
  \left[{I_{Nn}} + c\mathcal{L} \otimes G(s)K \right] x(s) = ({I_N} \otimes G(s))f(s) \hfill \\
  \Rightarrow \left[{I_{Nn}} + c V \Lambda V^{-1} \otimes G(s)K \right] x(s) = ({I_N} \otimes G(s))f(s) \hfill \\ 
  \Rightarrow (V \otimes I_n)\left[{I_{Nn}} + c \Lambda \otimes G(s)K \right](V^{-1} \otimes I_n) x(s) \hfill \\
  = ({I_N} \otimes G(s))f(s) \hfill \\ 
\end{gathered}\,\,.    
\end{equation}

Pre-multiplying \eqref{eq:42new} by $(V^{-1} \otimes I_n)$  and using  \(\hat x(s) = ({V^{ - 1} \otimes I_n})x(s)\) in \eqref{eq:42new} yields
\begin{equation}\label{eq:45}
\left[ {I_{Nn} + c\Lambda \otimes G(s)K} \right]\hat x(s) = (V^{-1} \otimes I_n)({I_N} \otimes G(s))f(s)\,,
\end{equation}
which yields
\begin{equation}\label{eq:45_new}
\hat x(s) = \left[ {I_{Nn} + c\Lambda \otimes G(s)K} \right]^{-1} [V^{-1} \otimes G(s)]f(s)\,.
\end{equation}

Assume $V = \left[ {{v_{ij}}} \right]$ and $V^{-1} = \left[ {{p_{ij}}} \right]$. Then, the state transformation yields
\begin{equation}\label{eq:49_state}
x(s) = (V \otimes I_n) \hat x(s) \Rightarrow {x_i}(s) = \sum\limits_{m = 1}^N {{v_{im}}{{\hat x}_m}(s)}\,.
\end{equation}

Equation \eqref{eq:45_new} is a block diagonal system and the size of each block is equal to the Jordan block corresponding with an eigenvalue \({\lambda _i}\) of \(\mathcal{L}\). Assume that all Jordan blocks in \(\Lambda \) are simple. Then, from \eqref{eq:45_new} for agent \(i\) one has  
\begin{equation}\label{eq:xhatnew}
{\hat x_i}(s) = \left[ {I_{n} + c\lambda_i G(s)K} \right]^{-1}G(s) \sum\limits_{j = 1}^N {{p_{ij}}{f_j}(s)}\,.
\end{equation}

Exploiting \eqref{eq:49_state} and \eqref{eq:xhatnew}, the state of agent $i$ can be represented by
\begin{equation}\label{eq:Statei}
{x_i}(s) = \sum\limits_{m = 1}^N {{v_{im}}\left[ {I_{n} + c\lambda_m G(s)K} \right]^{-1}G(s){\sum\limits_{j = 1}^N {p_{mj}}{f_j}(s)} }\,.
\end{equation}

The first eigenvalue of \(\mathcal{L}\) is zero and its corresponding right eigenvector is $1_N$, i.e., $v_{i1}=1\,\forall i=1,\dots,N$. Using this fact, \eqref{eq:Statei} turns into
\begin{equation}\label{eq:51}
\begin{gathered} 
{x_i}(s) = G(s)\sum\limits_{j = 1}^N {p_{1j}}{f_j}(s)  \hfill \\
+\sum\limits_{m = 2}^N {v_{im}}\left[ {I_{n} + c\lambda_m G(s)K} \right]^{-1}G(s){\sum\limits_{j = 1}^N {{p_{mj}}{f_j}(s)} } 
\end{gathered}\,\,.
\end{equation}

We now show that $\left[ {I_{n} + c\lambda_m G(s)K} \right]^{-1}$ is Hurwitz and, consequently, the second term of \eqref{eq:51} is bounded, regardless of the attack signal $f_j(s)$. To this end, we show that the poles of $\left[ {I_{n} + c\lambda_m G(s)K} \right]^{-1}$, i.e., the roots of $\left[\det \left( s{{I}_{n}}-A \right)+c{{\lambda }_{m}} adj(s{{I}_{n}}-A)BK\right]$, are identical to the roots of the characteristic polynomial $A-c\lambda_mBK$. That is \cite{Harville1997}
\begin{align*}
\begin{gathered}
  \det \left( s{{I}_{n}}-(A-c{{\lambda }_{m}}BK) \right)=\det (s{{I}_{n}}-A+c{{\lambda }_{m}}BK)= \hfill \\ 
  \det \left( s{{I}_{n}}-A \right) \det \left( I_n+c{{\lambda }_{m}}{{(s{{I}_{n}}-A)}^{-1}}BK \right) \hfill \\ 
 \end{gathered}\,\,, 
\end{align*}
which by using \eqref{eq:40}, implies that
\begin{equation}\label{eq:tm}
\begin{gathered}
  \det (s{I_n} - A + c{\lambda _m}BK) =  \hfill \\
  \det \left( s{{I}_{n}}-A \right) \det \left( I_n+c{{\lambda }_{m}}G(s)K \right)\\ 
\end{gathered}\,\,. 
\end{equation}

Equation \eqref{eq:tm} shows that the eigenvalues of $A-c\lambda_mBK$ are identical to the poles of $\left[ {I_{n} + c\lambda_m G(s)K} \right]^{-1}$. Since $A-c\lambda_mBK$ $\forall m=2,\dots,N$ are Hurwitz, therefore, $\left[ {I_{n} + c\lambda_m G(s)K} \right]^{-1}$ is also Hurwitz. Thus, the second term is bounded and has no contribution in destabilizing the system dynamics.

Consider the transfer function for agent dynamics \eqref{eq:1}, and the Laplace transform of the attacker dynamics \eqref{eq:12} as
$$G(s)=\frac{adj\left( sI-A \right)B}{\det \left( sI-A \right)},\quad f_j(s)=\frac{adj\left( sI-R \right)}{\det \left( sI-R \right)}f_j(0)\,,$$

\noindent
where $f_j(0)$ is the initial value of the attacker.

Based on Lemma 1, $p_{1j}$ in \eqref{eq:51} is zero for non-root nodes. Therefore, condition $\sum\nolimits_{j = 1}^N {{p_{1j}}{f_j}}\ne 0$ can only be satisfied if the attack is launched on root nodes. Assume that the attack is on root nodes and the attack dynamic that generates ${f_j}(s)$, i.e., the attack signal that is injected into a root node, has at least one common eigenvalue with the agent dynamics, $\lambda_{A_{k}}$. Then, \eqref{eq:51} can be expressed as
\begin{equation}\label{eq:newver}
\begin{gathered}
  {x_i}(s) = \sum\limits_{j = 1}^N {{p_{1j}}\frac{{\left[ {adj(sI - A)} \right]B\left[ {adj(sI - R)} \right]{f_j}(0)}}{{{{\left( {{s^2} + \lambda _{{A_k}}^2} \right)}^2}\left[ {\mathop \prod \limits_{i = 1,i \ne k}^n \left( {{s^2} + \lambda _{{A_i}}^2} \right)\left( {{s^2} + \lambda _{{R_i}}^2} \right)} \right]}}} \hfill \\
   + \sum\limits_{m = 2}^N {{v_{im}}} \left[ {I_{n} + c\lambda_m G(s)K} \right]^{-1}G(s)\sum\limits_{j = 1}^N {{p_{mj}}{f_j}(s)} \hfill \\
   \quad\quad \quad \quad  \quad \quad i = 1,\dots,N  \hfill \\
\end{gathered}\,\,.
\end{equation}

Since $\lambda_{A_{k}}$ is on imaginary axis and has multiplicity greater than 1 in \eqref{eq:newver}, it results that in time domain \eqref{eq:newver} converges to infinity as $t \to \infty$ \cite{dukkipati2005control}. Therefore, this type of attack on a root node destabilizes the entire network. This proves part 1.

For the proof of part 2, if $\sum\nolimits_{j = 1}^N {{p_{1j}}{f_j}}=0$, Then, \eqref{eq:51} becomes
\begin{equation}\label{eq:52lq}
{x_i}(s) = \sum\limits_{m = 2}^N {v_{im}}\left[ {I_{n} + c\lambda_m G(s)K} \right]^{-1}G(s){\sum\limits_{j = 1}^N {{p_{mj}}{f_j}(s)} }\,.
\end{equation}

As shown above, $\left[ {I_{n} + c\lambda_m G(s)K} \right]^{-1}$ is Hurwitz and thus $x_i(s)$ is bounded, regardless of the attack. However, the agents that are in the path of the attacker deviate from the consensus value by the amount of the attack signal $f_j(s)$.

On the other hand, if there is no common eigenvalue between system and attacker dynamics, i.e., ${E_R} \cap {E_A} =\emptyset$, the multiplicity of poles located on the imaginary axis is 1 and therefore, based on \eqref{eq:51}, the state of agents remains bounded. However, agents do not achieve consensus on the desired value because of the attack's effect. This completes the proof.\hfill $\blacksquare$ 
\end{pf} 

\noindent
\textbf{Remark 2.} Consider a NCS under attack represented by
\begin{align*}
{{\dot x}_i} = A{x_i} + Bu_i+B_ff_i\,.    
\end{align*}
Assume that the control protocol is given by
\begin{align*}
u = Kx\,,   
\end{align*}
with $x = {\left[ {x_1^T, \ldots ,x_N^T} \right]^T}$ and $u = {\left[ {u_1^T, \ldots ,u_N^T} \right]^T}$.
The closed-loop form, ignoring delays and packet losses, has the form
\begin{align*}
{{\dot x}} = A_{cl}{x} + B_{cl}{f}\,.    
\end{align*}

The system response can be defined by
\begin{align*}
x\left( t \right) = {e^{{A_{cl}}t}}x\left( 0 \right) + \int_0^t {{e^{{A_{cl}}\left( {t - \tau } \right)}}\left( {{B_{cl}}f} \right)d\tau }\,,
\end{align*}

Since the closed-loop system $A_{cl}$ is stable, the system response cannot be unbounded for a bounded attack signal. This is in contrast to the synchronization in the distributed MAS \eqref{eq:1} in which $A$ needs to be marginally stable, otherwise agent’s states converge to the origin.
\medskip

\subsection{Attack Analysis: A Graph-Theoretic Approach}
In this subsection, we use a graph theoretical approach to analyze the effect of attacks. These results comply with the results of Theorem 2, but reveal extra facts.

Define the set of root nodes and the set of non-root nodes as
\begin{equation}\label{eq:SrSnra}
\begin{gathered}
  S_r=\{v_1,\dots,v_r\}  \hfill \\
  S_{nr}=\{v_{r+1},\dots,v_N\}  \hfill \\ 
\end{gathered}\,\,. 
\end{equation}

The graph Laplacian matrix $\mathcal{L}$ can be partitioned as 
\begin{equation}\label{eq:Lp}
\mathcal{L}=\left[ \begin{matrix}
   {{\mathcal{L}}_{r\times r}} & {{0}_{r\times nr}}  \\
   {{\mathcal{L}}_{nr\times r}} & {{\mathcal{L}}_{nr\times nr}}  \\
\end{matrix} \right]\,,
\end{equation}
where $r$ and $nr$ denote the number of root nodes and non-root nodes, respectively. The  ${{\mathcal{L}}_{r\times r}}$ and ${{\mathcal{L}}_{nr\times nr}}$ are Laplacian matrices corresponding to the subgraphs of root nodes and non-root nodes, respectively. 
\medskip

\noindent
\textbf{Lemma 2.}\cite{lewis2013cooperative} Let $\Delta$ be a diagonal matrix with at least one nonzero positive element and $L$ be the graph Laplacian matrix. Then, $(L+\Delta)$ is nonsingular. \hfill $\blacksquare$ 
\medskip

\noindent
\textbf{Lemma 3.} Consider the partitioned Laplacian matrix \eqref{eq:Lp}. Then, ${{\mathcal{L}}_{nr\times nr}}$ is nonsingular and ${{\mathcal{L}}_{r\times r}}$ is singular.

\begin{pf} We first show that there exists at least a direct incoming link from the set $S_r$ to the set $S_{nr}$ defined in \eqref{eq:SrSnra}. Assume by contradiction that there is no such link. Then, none of the nodes in $S_r$ have access to nodes in $S_{nr}$, which contradicts the fact that $S_r$ is the set of root nodes. This violates Assumption 1. Consequently, the subgraph ${{\mathcal{L}}_{nr\times nr}}$ captures the interaction between all elements of $S_{nr}$ as well as the incoming links from $S_r$ to $S_{nr}$. The former is a positive semi-definite graph Laplacian matrix $L$ and the latter can be captured by a diagonal matrix $\Delta$ with at least one nonzero positive element added to it. Therefore, based on Lemma 2, ${{\mathcal{L}}_{nr\times nr}}$ is nonsingular.

On the other hand, the subgraph of the root nodes is strongly connected. This is because there is a link from each root node to all other root nodes, by the definition of a root node. We first show that there is no incoming link from the set $S_{nr}$ to the set $S_r$. Assume by contradiction that there is such a link. Then, $v_i \in S_{nr}$ has a path to all nodes in $S_r$, and since the subgraph of $S_r$ is strongly connected, it implies that $v_i \in S_r$, which contradicts the assumption that $v_i \in S_{nr}$. Therefore, the subgraph of root nodes is strongly connected and standalone with no incoming link from other nodes. Thus, ${{\mathcal{L}}_{r\times r}}$ is singular\cite{wu2007synchronization}. \hfill $\blacksquare$
\end{pf}

Let $p = {\left[ {p_1,\ldots,p_r,p_{r+1}, \ldots ,p_N} \right]^T}$ be the elements of $w_1^T$, the left eigenvector of $\mathcal{L}$ associated to its zero eigenvalue, where $[p_1,\dots,p_r]$ denotes the elements of $w_1^T$ associated to the root nodes and $[p_{r+1},\dots,p_{N}]$ are those for the non-root nodes. 

Consider the MAS \eqref{eq:1} with the control protocol \eqref{eq:39} under the attack signal \eqref{eq:12}. Define $S_A(t)=\{e^{\lambda_{A_1}t},\dots,e^{\lambda_{A_n}t} \}$ and $S_R(t)=\{e^{\lambda_{R_1}t},\dots,e^{\lambda_{R_m}t} \}$ as the sets of natural modes of the agent dynamic $A$ and the attacker dynamic $R$, respectively. Define the global state vector $x = {\left[ \bar{x}_r^T,\bar{x}_n^T \right]^T}$, where $\bar{x}_r = [x_1^T,\dots,x_r^T]^T$ and $\bar{x}_{n} = [x_{r+1}^T,\dots,x_N^T]^T$ denote the global vectors of the state of root nodes and the state of non-root nodes, respectively. Define the global attack vector $f = {\left[ {\bar{f}}_r^T, {\bar{f}}_{n}^T \right]^T}$, where ${\bar{f}}_r$ and ${\bar{f}}_{n}$ are the global vectors of attacks on root nodes and non-root nodes as 
\begin{equation}\label{eq:frfnr}
\begin{gathered}
  {\bar{f}}_r= [f_1^T,\dots,f_r^T]^T  \hfill \\
  {\bar{f}}_{n} = [f_{r+1}^T,\dots,f_N^T]^T \hfill \\ 
\end{gathered}\,\,, 
\end{equation}

Then, the global dynamics for \eqref{eq:1} and global form of the control protocol \eqref{eq:39} can be taken in the form
\begin{equation}\label{eq:g_lm}
  {\dot x} =(I_N \otimes A)x +(I_N \otimes B)u \,,
\end{equation}
\begin{equation}\label{eq:uglobal_lm}
  u = -(c\mathcal{L} \otimes K)x+f \,,
\end{equation}

In the absence of attacks, i.e., $f=0$, a nonzero control input $u$ in \eqref{eq:uglobal_lm} indicates a disagreement among agents and their neighbors. This disagreement eventually goes to zero, i.e., $u \to 0$, indicating that agents reach consensus. That is as $t\to \infty$, ${\dot x} \to (I_N \otimes A)x$. We call this the steady state of agents.

In the presence of attacks, the following results show that when $u$ in \eqref{eq:uglobal_lm} goes to zero, i.e., $-(c\mathcal{L} \otimes K)x+f \to 0$, although agents reach a steady state, i.e., ${\dot x} \to (I_N \otimes A)x$, they do not reach consensus. On the other hand, when $u$ in \eqref{eq:uglobal_lm} does not converge to zero, i.e., $-(c\mathcal{L} \otimes K)x+f \not \to 0$, conditions on which the attacker can destabilize the entire network or result in non-emergent behavior of agents are provided. To this end, we show that if $f \not \in \operatorname{Im} ({c{\mathcal{L}}} \otimes K)$ the network never reaches any steady state, i.e., ${\dot x} \not \to (I_N \otimes A)x$. It is also shown that if ${E_R} \subseteq {E_A}$, then, for a non-root node, $f \in \operatorname{Im} ({c{\mathcal{L}}} \otimes K)$ and for a root node, $f \in \operatorname{Im} ({c{\mathcal{L}}} \otimes K)$ if $\sum\nolimits_{k = 1}^r {{f_k} = 0}$, otherwise $f \not \in \operatorname{Im} ({c{\mathcal{L}}} \otimes K)$. We also show that if $f \not \in \operatorname{Im} ({c{\mathcal{L}}} \otimes K)$ and ${E_R} \cap {E_A} \ne \emptyset $, then the network becomes unstable.
\medskip

\noindent
\textbf{Lemma 4.} Consider the global dynamics of agents \eqref{eq:g_lm} with the control protocol \eqref{eq:uglobal_lm}. Agents reach a steady state, i.e., ${\dot x} \to (I_N \otimes A)x$, as $t \to \infty$, if and only if $f \in \operatorname{Im} (c\mathcal{L} \otimes K)$.
\begin{pf}
The global dynamic of agents \eqref{eq:g_lm} indicates that agents reach a steady state, if and only if, $-(c\mathcal{L} \otimes K)x+f \to 0$. This condition is satisfied, if and only if, there exists a bounded $x$ such that $-(c\mathcal{L} \otimes K)x+f=0$, i.e., $f \in \operatorname{Im} ({c{\mathcal{L}}} \otimes K)$. This completes the proof. \hfill $\blacksquare$ 
\end{pf}

\noindent
\textbf{Lemma 5.} Consider the MAS \eqref{eq:1} with global dynamics \eqref{eq:g_lm}. Assume that ${E_R} \not\subset {E_A}$. Then, $f \notin \operatorname{Im} (c\mathcal{L} \otimes K)$.

\begin{pf}
We prove this by contradiction. Let ${E_R} \not\subset {E_A}$ but $f \in \operatorname{Im} (c\mathcal{L} \otimes K)$. If $f \in \operatorname{Im} (c\mathcal{L} \otimes K)$, it implies that there exists a nonzero bounded $x$ such that $-(c\mathcal{L} \otimes K)x+f \to 0$. This concludes that the control signal $u$ in \eqref{eq:uglobal_lm} converges to zero, which results in $\dot x \to (I_N \otimes A)x$, and therefore, $\dot x_i \to Ax_i$. Using the modal decomposition, one has 
\begin{equation}\label{eq:T61_ss2}
{x_i}(t) \to \sum\limits_{j = 1}^n {({r_j}{x_i}(0)){e^{{\lambda_{A_j}}t}}{m_j}}\,,
\end{equation}
where $r_j$ and $m_j$ denote the left and right eigenvector associated with the eigenvalue ${\lambda_{A_j}}$ of the system dynamic $A$, respectively. On the other hand, $-(c\mathcal{L} \otimes K)x+f \to 0$ implies that 
\begin{equation}\label{eq:L6_Eqq1}
cK\sum\limits_{j \in {N_i}} {{e_{ij}}({x_j} - {x_i})} \to -{f_i}\,.
\end{equation}

The right-hand side of \eqref{eq:L6_Eqq1} is generated by the natural modes of the system dynamic $A$ whereas the left-hand side is generated by the natural modes of the attacker dynamic $R$. Based on our assumption, ${E_R} \not\subset {E_A}$ which implies that the attacker natural modes are different that the system natural modes. Therefore, \eqref{eq:L6_Eqq1} cannot be satisfied and thus, $f \notin \operatorname{Im} (c\mathcal{L} \otimes K)$, which contradicts the assumption. \hfill $\blacksquare$ 
\end{pf}

\noindent
\textbf{Lemma 6.} Consider the global dynamics \eqref{eq:g_lm} and assume that the attack signal is only injected into non-root nodes, i.e., $\bar{f}_r=0$, with $\bar{f}_r$ defined in \eqref{eq:frfnr}. If ${E_R} \subseteq {E_A}$, then, $f\in \operatorname{Im} (c\mathcal{L} \otimes K)$.

\begin{pf}
Note that, $f \in \operatorname{Im}(c\mathcal{L} \otimes K)$ if there exists a nonzero vector $x_{s}$ such that
\begin{equation}\label{eq:new1Lp}
(c\mathcal{L} \otimes K){x}_{s}=f\,,
\end{equation}
where $x_{s}$ can be represented as the global steady state solution of \eqref{eq:g_lm}. 
If \eqref{eq:new1Lp} holds, then $u=0$ and $\dot x_s = (I_N \otimes A)x_s$. This implies that ${x_i} \in span({S_A})$.

Define ${x}_{s}=[\bar{x}_{rs},\bar{x}_{ns}]^T$, where $\bar{x}_{rs}$ and $\bar{x}_{ns}$ are the global steady states of root nodes and non-root nodes, respectively.

Using \eqref{eq:Lp}, \eqref{eq:new1Lp} becomes
\begin{equation}\label{eq:lnr} 
 \left[ {\begin{array}{*{20}{c}}
  {c{\mathcal{L}_{r \times r}} \otimes K}&{{0_{r \times nr}}} \\ 
  {c{\mathcal{L}_{nr \times r}} \otimes K}&{c{\mathcal{L}_{nr \times nr}} \otimes K} 
\end{array}} \right]\left[ {\begin{array}{*{20}{c}}
  {\bar{x}_{rs}} \\ 
  {\bar{x}_{ns}} 
\end{array}} \right] = \left[ {\begin{array}{*{20}{c}}
  0 \\ 
  {{{\bar{f}}_{n}}} 
\end{array}} \right]\,,
\end{equation}
or, equivalently
\begin{equation}\label{eq:L5_lnr1} 
\left\{\begin{gathered}
  (c{\mathcal{L}_{r \times r}} \otimes K)\bar{x}_{rs} = 0 \hfill \\
  (c{\mathcal{L}_{nr \times r}} \otimes K)\bar{x}_{rs} + (c{\mathcal{L}_{nr \times nr}} \otimes K)\bar{x}_{ns} = {{\bar{f}}_{n}} \hfill \\ 
\end{gathered}\right.\,.
\end{equation}

As stated in Lemma 3, ${{\mathcal{L}}_{r\times r}}$ is singular with zero as an eigenvalue and ${{\mathbf{1}}_r}$, its corresponding eigenvector, and ${{\mathcal{L}}_{nr\times nr}}$ is nonsingular. Consequently, the solutions to \eqref{eq:L5_lnr1} can be written as
\begin{equation}\label{eq:L5_lnr2} 
\left\{ \begin{gathered}
  {\bar{x}_{rs}} = c_1{{\mathbf{1}}_r} \hfill \\
  {\bar{x}_{ns}} =   {(c{\mathcal{L}_{nr \times nr}} \otimes K)^{ - 1}}\left[-{(c{\mathcal{L}_{nr \times r}} \otimes K) c_1{{\mathbf{1}}_r} + {{\bar{f}}_{n}}} \right] \hfill \\ 
\end{gathered}  \right.\,,
\end{equation}
for some positive $c_1$ and $c$. Equation \eqref{eq:L5_lnr2} shows that the steady state value of the non-root nodes is affected by the attack signal value. If ${E_R} \not \subset {E_A}$, it results that ${\bar{x}_{ns}} \in span({S_A},{S_R})$ which contradicts ${x_i} \in span({S_A})$. Therefore, condition ${E_R} \subseteq {E_A}$ is necessary to conclude that for any $f=[0,\bar{f}_{n}]^T$ there exists a solution $x_{s}$ in the form of \eqref{eq:L5_lnr2} such that \eqref{eq:new1Lp} holds, which implies that $f\in \operatorname{Im} (c\mathcal{L} \otimes K)$. The proof is completed.\hfill $\blacksquare$ \end{pf}

\begin{thm}
Consider the MAS \eqref{eq:1} with the control protocol under attack \eqref{eq:39}. Assume that ${E_R} \subseteq {E_A}$. Then, $f\in \operatorname{Im}({{c\mathcal{L}}} \otimes K)$ if and only if 
\begin{equation}\label{eq:T3_eq1} 
\sum\limits_{k = 1}^N p_k{{f_k}}  = 0\,,
\end{equation}
where $p_k \, \forall k=1,\dots,N$ denote the elements of the left eigenvector of $\mathcal{L}$ corresponding with its zero eigenvalue.
\end{thm}
\begin{pf}
We first prove the necessary condition. It was shown in Lemma 6 that if ${E_R} \subseteq {E_A}$, then, $f = {[{ 0,{\bar{f}}_{n}}]^T} \in \operatorname{Im}( {c{{\mathcal{L}}} \otimes K} )$ regardless of ${\bar{f}}_{n}$. Therefore, whether $f \in \operatorname{Im}( {c{{\mathcal{L}}} \otimes K} )$ or $f \notin \operatorname{Im}( {c{{\mathcal{L}}} \otimes K})$ depends only on ${\bar{f}}_{r}$. If $f \in \operatorname{Im}( {c{{\mathcal{L}}} \otimes K})$, then, using \eqref{eq:Lp}, there exists a nonzero vector $\bar{x}_{rs}$ for root nodes such that
\begin{equation}\label{eq:lrr} 
(c{{\mathcal{L}}_{r\times r}} \otimes K) \bar{x}_{rs}={\bar{f}}_r\,,
\end{equation}
where $\bar{x}_{rs}$ can be considered as the global steady state of the root nodes. Moreover, equation \eqref{eq:lrr} holds, if ${E_R} \subseteq {E_A}$. Otherwise, based on Lemma 5, $f \not \in \operatorname{Im}({{c\mathcal{L}}} \otimes K)$. As stated in Lemma 3, ${{\mathcal{L}}_{r\times r}}$ is strongly connected and singular. Therefore, equation \eqref{eq:lrr} has a solution when ${\bar{f}}_r$ is in the column space of $(c{{\mathcal{L}}_{r\times r}} \otimes K)$. Since ${{\mathcal{L}}_{r\times r}}$ is strongly connected, the row sums of $\mathcal{L}_{r\times r}$ are all zero. Therefore, zero is one of its eigenvalues with $\bar{w}_1=[p_1,\dots,p_r]^T$, its corresponding left eigenvector. Pre-multiplying \eqref{eq:lrr} by $\bar{w}_1^T$ and using the fact that $\bar{w}_1^T{{\mathcal{L}}_{r\times r}}=0$, one has
\begin{equation}\label{eq:T3_lrr} 
w_1^T(c{\mathcal{L}_{r \times r}} \otimes K){\bar x_{rs}} = w_1^T{\bar f_r} = 0 \Rightarrow \sum\limits_{k = 1}^r {{p_k}} {f_k} = 0\,.
\end{equation}

As stated in Lemma 1, $p_i=0$ for non-root nodes. Therefore, \eqref{eq:T3_lrr} can be written as $\sum\nolimits_{k = 1}^N p_k{{f_k}}  = 0$.

Now, we prove the sufficient condition by contradiction. Consider $f \in \operatorname{Im}( {c{{\mathcal{L}}} \otimes K})$ but condition \eqref{eq:T3_eq1} is not satisfied, i.e., $\sum\nolimits_{k = 1}^N p_k{{f_k}} \ne 0$. Note that, $f \in \operatorname{Im}( {c{{\mathcal{L}}} \otimes K})$ implies that there exists a nonzero vector $\bar{x}_{rs}$ such that \eqref{eq:lrr} holds. Pre-multiplying \eqref{eq:lrr} by $\bar{w}_1^T$ results in $w_1^T(c{\mathcal{L}_{r \times r}} \otimes K){\bar x_{rs}} = w_1^T{\bar f_r} = 0$ which holds if $\sum\nolimits_{k = 1}^r p_k{{f_k}} = 0$. This contradicts the assumption. Therefore, $f \in \operatorname{Im}( {c{{\mathcal{L}}} \otimes K})$, if and only if condition \eqref{eq:T3_eq1} holds. \hfill $\blacksquare$  
\end{pf}

\begin{thm}
Consider the MAS \eqref{eq:1} with the control protocol under attack \eqref{eq:39}. If the attack is on root nodes, then,
\begin{enumerate}
\item All agents remain stable, but the network shows no emergent behavior, if $f\in \operatorname{Im}(c\mathcal{L} \otimes K)$.
\item The entire network is destabilized when ${E_R} \cap {E_A} \ne \emptyset$ and $f\notin \operatorname{Im}(c\mathcal{L} \otimes K)$. 
\end{enumerate}
\end{thm}
\begin{pf}
The global dynamics \eqref{eq:g_lm} can be written as
\begin{equation}\label{eq:T5_GSU1}
\dot x = ({I_N} \otimes A)x + ({I_N} \otimes B)\left[ {-(c\mathcal{L} \otimes K)x + f} \right]\,.
\end{equation}

One can see from the dynamics that if the second term in \eqref{eq:T5_GSU1} converges to zero, i.e., if $f \in \operatorname{Im}(c\mathcal{L} \otimes K)$, then, as $t\to \infty$, one has for the the agent dynamics $\dot x_i \to Ax_i$, which indicates their stability. On the other hand, $f \in \operatorname{Im}(c\mathcal{L} \otimes K)$ implies that $cK\sum\nolimits_{j \in {N_i}} {{e_{ij}}({x_j} - {x_i})}  + {f_i} \to 0$ and then, ${x_i} \to \frac{1}{{{d_i}}}\sum\nolimits_{j \in {N_i}} {{e_{ij}}{x_j}}  + \frac{{{{({K^T}K)}^{ - 1}}K}}{{c{d_i}}}{f_i}$. Therefore, the state of the agent under direct attack is deviated from the state of its neighbors with a value proportional to $f_i$. Using Theorem 1, the agents that have a path to the compromised agents also  deviate from their real valued state and, therefore, the entire network shows non-emergent behavior.

To prove Part 2, assume that there exists at least one common mode between system dynamic $A$ defined in \eqref{eq:1} and attacker dynamic $R$ defined in \eqref{eq:12}, i.e., ${E_R} \cap {E_A} \ne \emptyset $. If the attack signal $f$ does not go away eventually from \eqref{eq:T5_GSU1}, the entire network becomes unstable. This is because the multiplicity of eigenvalues located on the imaginary axis would be greater than 1. This situation happens, if and only if, $-(c\mathcal{L} \otimes K)x+f \not \to 0$, or equivalently, $f \notin \operatorname{Im}(c\mathcal{L} \otimes K)$. In this case, $f$ acts as an input to \eqref{eq:T5_GSU1}. Using Assumption 2 and considering $\lambda_{A_k} \in {E_R} \cap {E_A}$, conclude that the multiplicity of the common poles related to $\lambda_{A_k}$ located on the imaginary axis \cite{dukkipati2005control} are greater than 1. Therefore, it makes the MAS \eqref{eq:T5_GSU1} unstable. \hfill $\blacksquare$  
\end{pf}

\noindent
\textbf{Remark 3.} Attacks on sensors disrupt sensor measurements and, consequently, the control protocol that uses this measurement data generates erroneous commands for actuators. Moreover, attacks on the communication links and actuators also result in a wrong command generated by the control protocol. For agents with multiple sensors and/or actuators, if only one sensor or actuator of a root node is attacked and the attacker satisfies conditions of Theorems 2 and 4, the entire network becomes unstable.
\medskip

\noindent
\textbf{Remark 4.} The effects of an attacker on a network depend on the attack signal dynamics. As shown in Theorem 2, if an attacker does not have any knowledge of agent's dynamics, it cannot destabilize the network. However, the network shows an emergent misbehavior. On the other hand, as stated in Theorems 2 and 4, an attacker does not need to have the full knowledge of the network topology and the agent's dynamics to destabilize the entire network. Attack signal requires having at least one common eigenvalue with agent's dynamics to make the entire network unstable. To this end, an attacker can exploit the security of the network by eavesdropping and monitoring the transmitted data to identify at least one of eigenvalues of the agent dynamics, then launch a signal with the same frequency to a root node to destabilize the entire network.
\medskip

Define the local neighborhood tracking error for the MAS \eqref{eq:1} with the control protocol \eqref{eq:2} as
\begin{equation}\label{eq:L_Te}
{\varepsilon_i} = {\sum\limits_{j \in {N_i}} e_{ij}{\left( {{x_j} - {x_i}} \right)} },
\end{equation}

The global form of \eqref{eq:L_Te} becomes
\begin{equation}\label{eq:G_Te}
{\varepsilon} = -(\mathcal{L} \otimes I_n) x\,.
\end{equation}

Theorem \ref{ErEa} shows that despite the presence of an attacker, the local neighborhood tracking error \eqref{eq:L_Te} becomes zero for intact agents if $E_R \subseteq E_A$.
\medskip

\begin{thm}\label{ErEa}
Consider the MAS \eqref{eq:1} with the control input under attack \eqref{eq:39}. Let the non-root node $i$ be under attack. Then,
\begin{enumerate}
    \item If $E_R \subseteq E_A$, then the local neighborhood tracking error converges to zero for all intact agents.
    \item If $E_R \not \subset E_A$, the local neighborhood tracking error does not converge to zero for agents that have a path to a compromised agent.
\end{enumerate}

\end{thm}
\begin{pf}
It is shown in Lemma 6 that if $E_R \subseteq E_A$, then, $f\in \operatorname{Im}(c\mathcal{L} \otimes K)$. Therefore, using \eqref{eq:G_Te} one has
\begin{equation}\label{eq:T5_eq1ep}
(c\mathcal{L} \otimes K)x=f \Rightarrow c(I_N \otimes K)\varepsilon=-f\,,
\end{equation}
or, equivalently
\begin{equation}\label{eq:T5_eq2ep}
ck\varepsilon_j=-f_j \quad \forall j=1,\dots,N\,.
\end{equation}

Since $f_j=0$ for intact agents, it implies that the local neighborhood tracking error is zero for intact agents. This proves Part 1. 

We now use the contradiction to prove Part 2. Assume that $E_R \not \subset E_A$ and $\varepsilon_j \to 0$ for all intact agents, which concludes $\dot x_j\to Ax_j$. This implies that $x_j$ is generated by the natural modes of the system dynamic $A$, i.e., $x_j \in span\{S_A\}$. However, for the compromised agent $i$, one has $\dot x_i=Ax_i+cBK\sum\nolimits_{j \in {N_i}} {{e_{ij}}({x_j} - {x_i})}+Bf_i$, which denotes that $x_i$ is composed of the natural modes of the system dynamic $A$ and the attacker dynamic $R$, i.e., $x_i \in span\{S_A,S_R\}$. On the other hand, $\varepsilon_j \to 0$ for the neighbors of the compromised agent implies that $\sum\nolimits_{j \in {N_i}} {{e_{ij}}({x_j} - {x_i})}\to 0$. Since the compromised agent does not have the same dynamic as other agents, $\varepsilon_j \to 0$ if and only if, $f_i$ has same dynamic with agents, i.e., $E_R \subseteq E_A$, which contradicts the assumption. Therefore, $\varepsilon_j \not\to 0$ for the compromised agent and the intact agents that have a path to it. \hfill $\blacksquare$
\end{pf}

The following uses Lyapunov method to show that for the special case of the single integrator MAS, the local neighborhood tracking error is zero for any constant attack. This complies with the results of Theorem 5.
\medskip

\noindent
\textbf{Lemma 7.} Consider the graph Laplacian matrix $\mathcal{L}$. Then, $\ker(\mathcal{L}+\mathcal{L}^T)= \emptyset $, if the graph is not balanced and $\ker(\mathcal{L}+\mathcal{L}^T)= span\{{{\mathbf{1}}_N}\}$, if it is balanced.

\begin{pf}
It is shown in \cite{Zhang2012LyapunovLewis} that
\begin{equation}\label{eq:L6_LP}
{x^T}(\mathcal{L} + {\mathcal{L}^T})x = \sum\limits_{i,j = 1}^N {{e_{ij}}{{({x_j} - {x_i})}^2}}\,.
\end{equation}

From [\cite{Zhang2012LyapunovLewis}, Lemma 8], one has $\ker (\mathcal{L} + {\mathcal{L}^T}) = \left\{ {\left. x \right|{x^T}(\mathcal{L} + {\mathcal{L}^T})x = 0} \right\}$. On the other hand, \eqref{eq:L6_LP} implies that ${x^T}(\mathcal{L} + {\mathcal{L}^T})x =0$, if and only if, $x_i=x_j, \, \forall i,j=1,\dots,N $. This indicates that $\ker(\mathcal{L}+\mathcal{L}^T) \subseteq span\{{{\mathbf{1}}_N}\}$. 

We now show that if the graph is not balanced, $x_i=x_j=0, \, \forall i,j=1,\dots,N $ is the only option and, therefore, $\ker(\mathcal{L}+\mathcal{L}^T)= \emptyset $. Assume that there exists a nonzero vector $x$ such that $(\mathcal{L} + {\mathcal{L}^T})x=0$. Since $\ker(\mathcal{L}+\mathcal{L}^T) \subseteq span\{{{\mathbf{1}}_N}\}$, therefore,  $x={{\mathbf{1}}_N}c$. That is,   
$(\mathcal{L} + {\mathcal{L}^T}){{\mathbf{1}}_N}c=0$ or equivalently, $ \mathcal{L}{{\mathbf{1}}_N}c + {\mathcal{L}^T}{{\mathbf{1}}_N}c=0$. The null space of the graph Laplacian matrix $\mathcal{L}$ is ${{\mathbf{1}}_N} $, i.e., $\mathcal{L}{{\mathbf{1}}_N}=0$, and this results in ${\mathcal{L}^T}{{\mathbf{1}}_N}c=0$, which is true if the graph is balanced. This contradicts the assumption that the graph is unbalanced. Therefore, $\ker(\mathcal{L}+\mathcal{L}^T)= \emptyset $. 

It is shown in [\cite{Zhang2012LyapunovLewis}, Lemma 9 and 11] that for the case of a balanced graph, $\ker(\mathcal{L}+\mathcal{L}^T) = span\{{{\mathbf{1}}_N}\}$. \hfill $\blacksquare$  
\end{pf}
\begin{thm}
Consider a MAS with single-integrator dynamics as
\begin{equation}\label{eq:MAS_SI}
{\dot x_i} = u_i, \quad i=1,\dots,N\,,
\end{equation}
and the control protocol \eqref{eq:39} under attack $f$. Then, for a constant attack, $(-\mathcal{L}x + f) \to 0 \,$ as $t \to \infty$.
\end{thm}
\begin{pf} Consider the Lyapunov function candidate for the MAS \eqref{eq:MAS_SI} as
\begin{equation}\label{eq:S_lyap}
V(x,f) = (-\mathcal{L}x + f)^T(-\mathcal{L}x + f)\,.
\end{equation}

The attack signal is constant, therefore, $\dot f=0$. Taking the time derivative of the Lyapunov function candidate \eqref{eq:S_lyap} yields
\begin{equation}\label{eq:Diff_lyap}
\begin{gathered} 
  \dot V(x,f) =\hfill \\
  {( - \mathcal{L}\dot x + \dot f)^T}( - \mathcal{L}x + f) + {( - \mathcal{L}x + f)^T}( - \mathcal{L}\dot x + \dot f) \hfill \\
  =  - {( - \mathcal{L}x + f)^T}({\mathcal{L}^T} + \mathcal{L})( - \mathcal{L}x + f) \leqslant 0 \,. \hfill \\ 
\end{gathered}
\end{equation}

By LaSalle's invariance principle \cite{isidori1995nonlinear}, the trajectories converge to the largest invariant set, $S= \{ (x,f)|\dot V(x,f)=0 \}$. Based on \eqref{eq:Diff_lyap}, $\dot V(x,f) = 0$ if
\begin{equation}\label{eq:T4_lyap1-1}
\left\{ \begin{gathered}
    \vspace{-3pt}
  ( - \mathcal{L}x + f) \in \ker ({\mathcal{L}^T} + \mathcal{L}) \hfill \\
    \vspace{-3pt}
  \quad \quad or \hfill \\
  ( - \mathcal{L}x + f) = 0 \hfill \\ 
\end{gathered}  \right.\,\,.
\end{equation}

As stated in Lemma 7, $\ker ({\mathcal{L}^T} + \mathcal{L})=\emptyset$ when the graph is not balanced. Therefore, the only solution to the \eqref{eq:T4_lyap1-1} is $-\mathcal{L}x+f=0$.

When the graph is balanced, based on Lemma 7, conditions \eqref{eq:T4_lyap1-1} become
\begin{equation}\label{eq:T4_lyap1-2}
\left\{ \begin{gathered}
   \vspace{-3pt}
   - \mathcal{L}x + f = c{{\mathbf{1}}_N} \hfill \\
   \vspace{-3pt}
   \quad \quad or \hfill \\
   - \mathcal{L}x + f = 0 \hfill \\ 
\end{gathered}  \right.\,\,.
\end{equation}

Consider that the attack is only on non-root nodes. The first equation of \eqref{eq:T4_lyap1-2} implies that for root node $i$, one has $\sum\nolimits_{j \in {N_i}} {({x_j} - {x_i})}  = c$ which results in $\dot x_i =c$ and denotes that the root nodes converge to infinity as $t \to \infty$. This violates the Theorem 1 because we showed in Lemma 5 that the subgraph of root nodes is strongly connected without any incoming link from non-root nodes and then, the attack on non-root nodes cannot affect the root nodes. Therefore, the only solution in this case is $c=0$.

For the case of constant attacks on root nodes, one has $\sum\nolimits_{j \in {N_i}} [{{x_j} - ({x_i}-f_i)}]  = c$. We showed in Theorem 2 that for this kind of attack, the entire network converges to infinity, i.e., $x_i=x_j \to \infty$. Therefore, $f_i$ and $c$ do not have any effects on the state of nodes. Moreover, multiplying both sides of the first equation in \eqref{eq:T4_lyap1-2} by $w_1^T$, the left eigenvector of the Laplacian matrix $\mathcal{L}$ corresponding to the zero eigenvalue, yields
\begin{equation}\label{eq:T4_lyap1-3}
- w_1^T\mathcal{L}x + w_1^Tf = cw_1^T{{\mathbf{1}}_N} \Rightarrow \sum\limits_{i = 1}^N p_i{{f_i}}  = c\,.  
\end{equation}

As shown in Theorem 3, $f \in \operatorname{Im}(\mathcal{L})$, if and only if, $\sum\nolimits_{i = 1}^N p_i{{f_i}}  = 0$, which concludes $c=0$. Therefore, the only solution to \eqref{eq:T4_lyap1-2} is $- \mathcal{L}x + f = 0$.\hfill $\blacksquare$
\end{pf}

\section{Discussion}
In this section, the adverse effects of attacks on root nodes and non-root nodes are thoroughly discussed based on the results from previous sections. 
\subsection{Attack on root nodes}
In this subsection, the adverse effects of attacks on root nodes are discussed. It is discussed on some real-world applications that, in case of a short-duration attack signal on a root node, the number of agents in the network has a significant impact on its performance. It is also shown what occurs if a root node is entirely compromised.
\medskip

\noindent
\textbf{Corollary 1.} Consider the MAS \eqref{eq:1} with the control protocol under attack \eqref{eq:39}. The entire network becomes unstable if 
\begin{enumerate}
    \item ${E_R} \subseteq {E_A}$
    \item $\sum\limits_{k = 1}^N {{p_k}{f_k} \ne 0}$
\end{enumerate}
\begin{pf}
Corollary 1 is the combination of Theorem 3 and Theorem 4. \hfill $\blacksquare$
\end{pf}

\noindent
\textbf{Remark 5.} An attack on only one root node makes the entire network unstable, if the attack signal dynamics has at least one eigenvalue in common with the agent's dynamic. This is because based on Theorem 4, attacks on root nodes destabilize the entire network if the attack signals on root nodes satisfies $\sum\nolimits_{i = 1}^N p_i{{f_i}} \ne 0$. For an attack on only one root node, $\sum\nolimits_{i = 1}^N p_i{{f_i}} = f_j \ne 0$, where $j$ is the attacked root node.
\medskip

Based on the results of Corollary 1 and Theorem 4, one can conclude that existing disturbance attenuation techniques, such as H$_\infty$ \cite{Jiao2016361}, fail to attenuate the adverse effects of the attacker on the network performance. This is because the goal in these approaches is to attenuate the effect of disturbance on the local neighborhood tracking error. In the presence of a stealthy attack, however, the state of all agents converge to infinity simultaneously, and consequently, the local neighborhood tracking error converges to zero eventually despite instability. On the other hand, existing mitigation techniques based on the discrepancy between the state of the agent and its neighbors \cite {LeBlanc2013resilient,Klotz2016Dixon,Pasqualetti2012TAC,Feng2017ACC} might not work for this type of attacks, as the agent’s state show the same misbehavior and grow unbounded all together and, therefore, it might be impossible to identify the compromised agent based on the discrepancy between an agent’s state and its neighbors states. For practical applications, since the state of all agents grows unbounded simultaneously, one may have to shut down the entire network once the states of agents exceed their permitted bounds. Therefore, the reliability of root nodes has a very crucial role in providing resiliency for the network, and one would be better to invest in making root nodes more secure. Novel resilient and secure control protocols are needed to be developed for these types of attacks.

In the case that the dynamics of the attack signal have no common eigenvalue with the system dynamics, as shown in Theorem 2 and Theorem 4, it cannot destabilize the network. However, the entire network shows a misbehavior. The local neighborhood tracking error is nonzero and, therefore, this type of attack can be mitigated to some extent by the H$\infty$ disturbance attenuation technique or game-based approach. However, the H$\infty$ technique is conservative and one needs to develop novel detection and mitigation techniques to repair and bring back the compromised agents to the network. 

The next example clarifies the adverse effects of the attack signal on root nodes.
\medskip

\begin{figure}[!ht]
\begin{center}
\includegraphics[width=2in,height=1.5in]{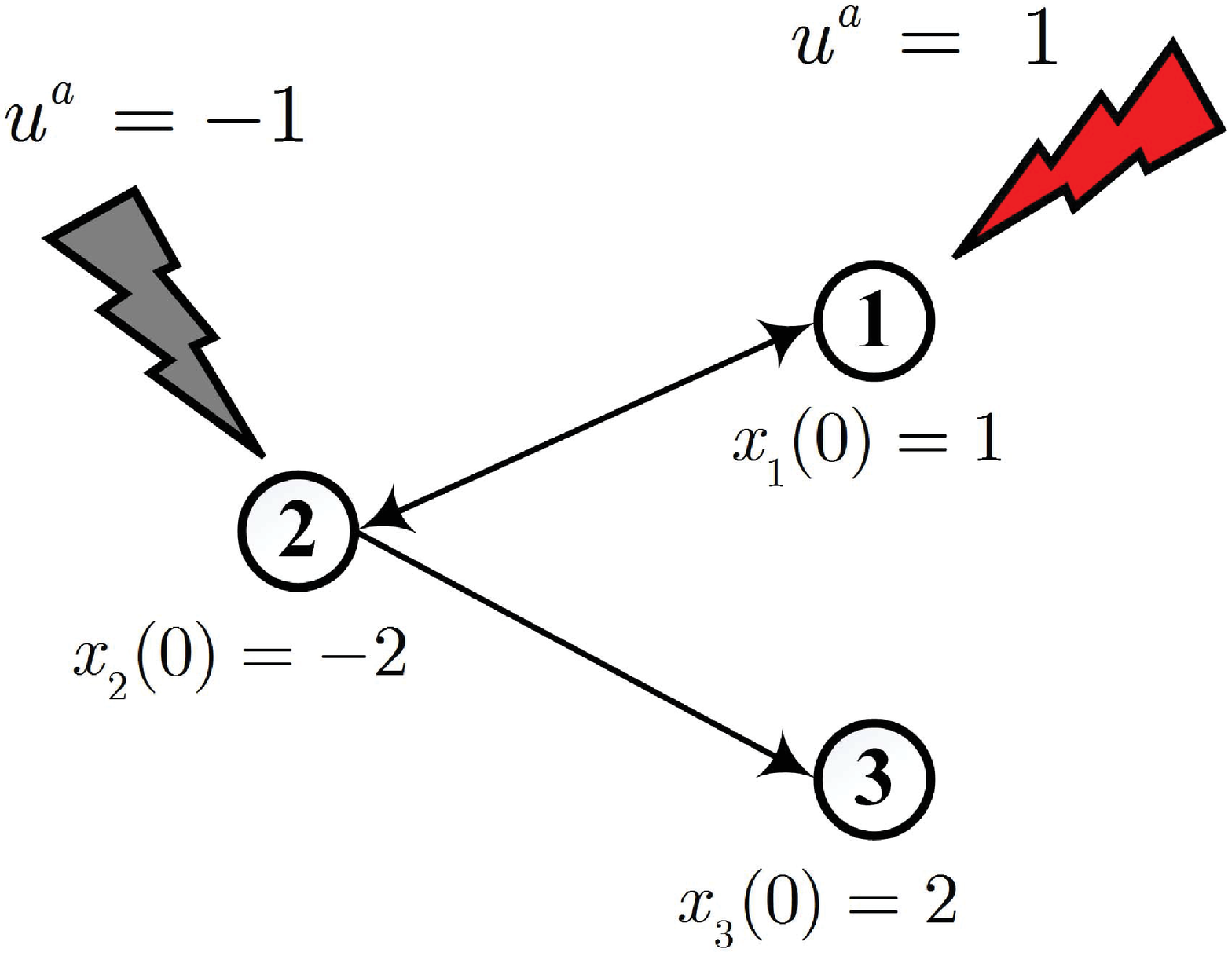}
\vspace{-5pt}\caption{Communication topology.}\label{fig:Fig1}
\captionsetup{justification=centering}
\end{center}
\end{figure}

\noindent
\textbf{Example 1.} Consider 3 agents communicating with each other according to the graph shown in Fig. \ref{fig:Fig1}.

The agents dynamics and control protocol are \eqref{eq:MAS_SI} and \eqref{eq:39}, respectively. When Agent 1 is attacked, the agents reach a steady state if
\begin{equation}\label{eq:AT_ag1_inf}
\begin{gathered}
  {{\dot x}_1}(t) =  - ({x_1} - {x_2})+1=0\Rightarrow (x_1-x_2)=1 \hfill\\
  {{\dot x}_2}(t) =  - ({x_2} - {x_1})=0\Rightarrow (x_1-x_2)=0 \hfill\\
  {{\dot x}_3}(t) =  - ({x_3} - {x_2})=0\Rightarrow (x_3-x_2)=0  \hfill\\ 
\end{gathered}\,\,.
\end{equation}

However, this needs both $(x_1-x_2)=1$ and $(x_1-x_2)=0$, which is impossible. Therefore, agents never reach a steady state and converge to infinity.

The system is single integrator and when an attacker forces a constant disrupted signal, i.e., $u^a=1$, into the root node 1, using Laplace transform, one has
\begin{equation}\label{eq:55}
{x_1}(s) = \frac{1}{s^2}+\frac{u(s)}{s}+\frac{x_1(0)}{s}\,,
\end{equation}
which implies that all agents tend to infinity. This confirms the results of Theorem 2.

Now, let Agents 1 and 2 as root nodes be under attack so that $\sum\nolimits_{k = 1}^2 p_k{{f_k}}=0$. Then, the steady state of agents can be written as
\begin{equation}\label{eq:AT_ag12}
\begin{gathered}
  {{\dot x}_1}(t) =  - ({x_1} - {x_2})+1=0\Rightarrow x_1=x_2+1\Rightarrow x_1 = 0 \hfill \\
  {{\dot x}_2}(t) =  - ({x_2} - {x_1})-1=0\Rightarrow x_2=x_1-1\Rightarrow x_2 = -1 \hfill\\
  {{\dot x}_3}(t) =  - ({x_3} - {x_2})=0\Rightarrow x_3 = x_2=0\Rightarrow x_3 = -1  \hfill\\ 
\end{gathered}
\end{equation}

It can be seen from \eqref{eq:AT_ag12} that although agents do not tend to infinity, they are disrupted from their desired value, which complies with the result of Theorem 2 and Theorem 3.

Note that although the results of Corollary 1 might require a consistent attack signal, a short-duration attack, caused by either an attack signal with a short-time duration or an attack signal that is detected and removed, can also result in a catastrophe. This type of attack on a root node has a permanent effect on the consensus value of the entire network. The attacker, once removed, implicitly changes the initial condition of the compromised agent. On the other hand, as stated in Lemma 3, the subgraph of root nodes are strongly connected and, therefore, based on Theorem 1, the effect of an attack signal on every root node propagates to all other root nodes. Therefore, an attack on a root node, even after it is removed, has permanent effect on the consensus value and the agents can agree on a wrong value. The attack signal can cause the agent's state to exceed their acceptable and safe values in its short-duration and ,therefore, the entire network might need to be shut down.

For this type of attack, the vulnerability of the network depends on the ratio of compromised root nodes to intact root nodes. For example, consider the demand-response management application in smart grids for which agents communicate to reach consensus on the aggregate power consumption value \cite{Chen2014DSM,Grammatico2015Lygeros,nasirian2014distributed}. In this application, there are possibly thousands of agents and each agent needs to be a root node to contribute to the aggregate power consumption value. Therefore, if a small portion of nodes are attacked, the attacker cannot deviate the aggregate value (for example the average value) from its target value considerably. For a network with a few number of agents, for example heading consensus of a few number of vehicle, if only one root node is attacked for a short period of time, it can significantly affect the consensus value.

\subsubsection{Fully compromised root node}

In the case that a root node is entirely compromised, it acts as an illegitimate leader. This is because it does not listen to the other root nodes and therefore, does not update its own information by the data received from its neighbors. As stated in Lemma 3, the subgraph of root nodes is strongly connected without any incoming link from non-root nodes. Therefore, this node is the only root node in the network. In such a situation, the compromised root node dictates its own information to the entire network. This concludes that the entire network is under the control of the attacker and agents converge to the state provided by the illegitimate leader.  
\subsection{Attack on non-root nodes}
In this subsection, the effects of attack signals on non-root nodes are discussed. 
\medskip

\noindent
\textbf{Corollary 2.} Attacks on non-root nodes, regardless of the number of compromised agents and energy of the attack signal, cannot destabilize the network.
\begin{pf}
As stated in Theorem 2 and Theorem 4, only attacks on root nodes can destabilize the entire network. \hfill $\blacksquare$
\end{pf}

\noindent
\textbf{Corollary 3.} Attacks on non-root nodes make the local neighborhood tracking error zero, when the condition of Part 1 of Theorem 5 holds.
\begin{pf}
The proof is similar to the proof of Theorem 5. \hfill $\blacksquare$
\end{pf}
The results of Corollary 3 indicates that existing H$_\infty$ attenuation techniques and game-based approaches cannot mitigate this type of attack. This is because the local neighborhood tracking error is zero despite attack and this can mislead the disturbance attenuation techniques. 

In case that the attack signal has a short-duration, as long as the attacker injects its adverse signal into the compromised agent, agents that have a path to it cannot synchronize. However, when it is removed, all agents will roll back to the desired consensus value provided by root nodes. Despite, the effects of the attacker on the network performance for a short period of time cannot be neglected and it can cause a catastrophe. 

The following example interprets the effects of an attacker on non-root nodes.
\medskip

\begin{figure}[!ht]
\begin{center}
\includegraphics[width=2in,height=1.5in]{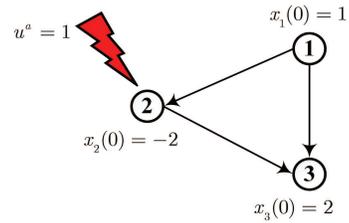}
\vspace{-5pt}\caption{Communication topology.}\label{fig:Fig2}
\captionsetup{justification=centering}
\end{center}
\end{figure}

\noindent
\textbf{Example 2.} Consider 3 agents communicating with each other according to the graph depicted in Fig. \ref{fig:Fig2}.

Assume agents with dynamics and control protocol \eqref{eq:MAS_SI} and \eqref{eq:39}, respectively.

Without any attack, the agents synchronize to the desired value. In this case, it is the initial value of Agent 1. When Agent 2 is attacked by injecting a constant and persistent signal to its actuators as $u^a$, then according to Theorem 1, Agents 2 and 3 do not reach consensus and, in the steady state, one has
\begin{equation}\label{eq:37}
\begin{array}{l}{{{\dot{x}}}_{1}}={{u}_{1}}=0\Rightarrow {{x}_{1}}=1\\ {{{\dot{x}}}_{2}}={{u}_{2}}=-\left( {{x}_{2}}-{{x}_{1}}\right)+u^a=0\Rightarrow {{x}_{2}}={{x}_{1}}+u^a=2\\ {{{\dot{x}}}_{3}}={{u}_{3}}=-\left( {{x}_{3}}-{{x}_{2}} \right)-\left(x_3-x_1\right)=0\Rightarrow x_3=\frac{{{x}_{1}}+{{x}_{2}}}{2}=1.5\end{array}\,\,.
\end{equation}

This confirms that attacks on a non-root node only affect other non-root nodes in the path of the compromised agents. The local neighborhood tracking error for Agents 2 and 3 is
\begin{equation}\label{eq:37_TRE}
\begin{array}{*{20}{l}}
  {{\varepsilon _2} = {x_2} - {x_1} = 1} \\ 
  {{\varepsilon _3} = {x_3} - {x_1} + {x_3} - {x_2} = 0} 
\end{array}\,,
\end{equation}
which complies with Theorem 5 that the local neighborhood tracking error is zero for all intact agents except compromised agents.

Therefore, even though local neighborhood tracking error is zero for intact agents, they do not converge to the desired consensus value. 

\section{Simulation Results}
In this section, a simulation example is provided to confirm our main results. Attack on both root nodes and non-root nodes are discussed.

\begin{figure}[!ht]
\begin{center}
\includegraphics[width=1.4in,height=1.4in]{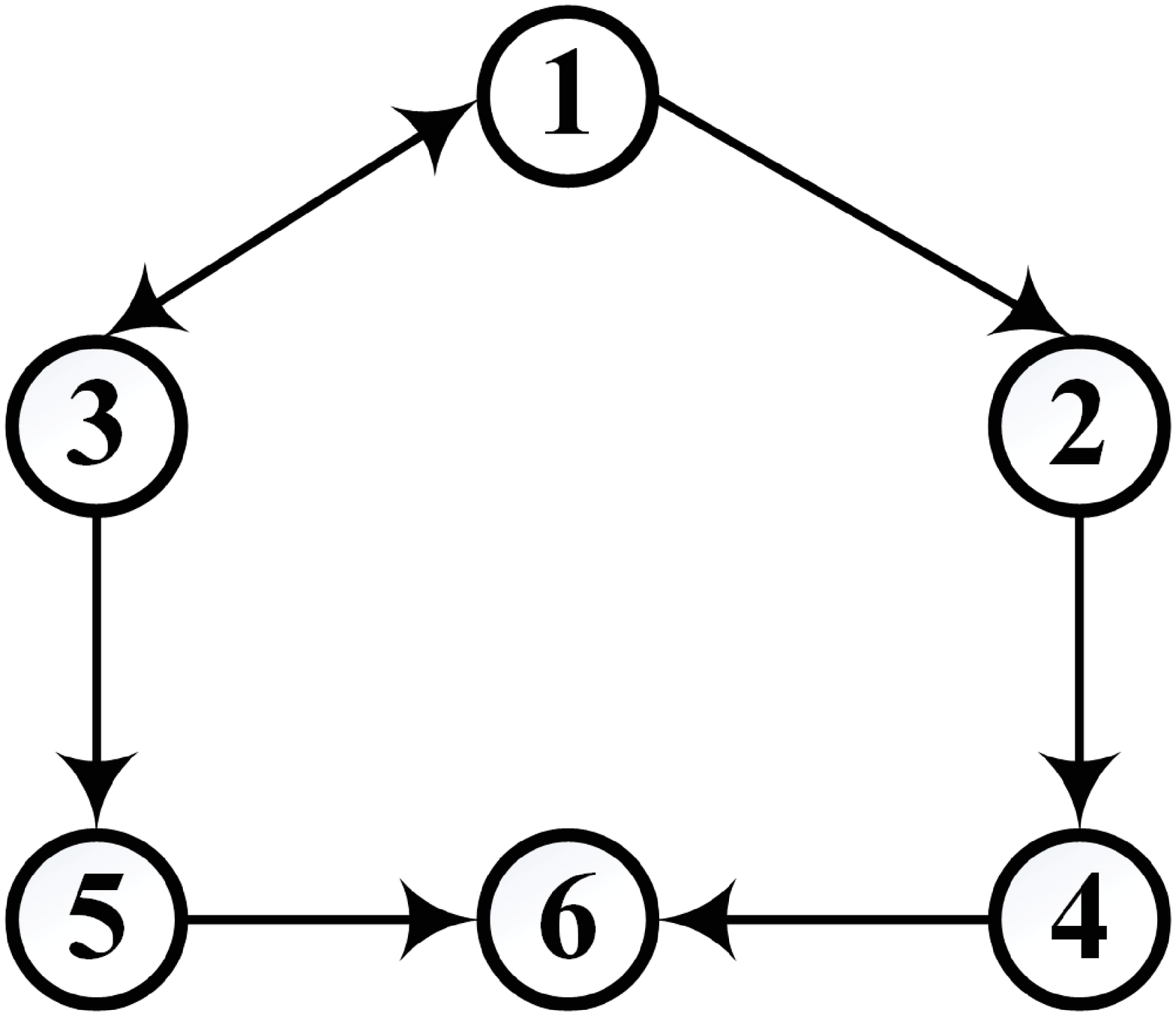}
\vspace{0pt}
\caption{Communication topology.}\label{fig:CT}
\captionsetup{justification=centering}
\end{center}
\end{figure}

Consider 6 agents communicating with each other by graph topology shown in Fig. \ref{fig:CT}. 

The dynamics of the agents are given by 
\begin{equation}\label{SR1}
\begin{gathered}  {{\dot x}_i} = \left[ {\begin{array}{*{20}{c}}  0&-1 \\
  { 1}&0 \end{array}} \right]{x_i} + \left[ {\begin{array}{*{20}{c}}  1 \\   0 
\end{array}} \right]{u_i} \hfill \\  \quad \quad \quad i = 1,\dots,6 \hfill \\ 
\end{gathered}\,\,,
\end{equation}
where $u_i$ is defined in \eqref{eq:2}. For simulation, Three types of attacks are considered as follows
\begin{equation}\label{eq:57}
\left\{ \begin{array}{ll}
         f_1 = \sin(t) & \mbox{$t\ge20$}\\
         f_2 = \sin(t) & \mbox{$20\le t<35$}\\
         f_3 = \sin(10t) & \mbox{$t\ge20$}\\
         \end{array} \right.
\end{equation}
where $f_1$ and $f_2$ denote the long-duration and short-duration attack signals with common eigenvalue with the agent's dynamics and $f_3$ is the long-duration attack signal without common eigenvalue with agent's dynamic.

\subsection{Attack on Root Node}
In this subsection, the effects of the attack signals on root nodes are shown.

\begin{figure}
    \centering
    \begin{subfigure}[b]{0.23\textwidth}
        \centering
        \includegraphics[width=1\linewidth,height=3.2cm]{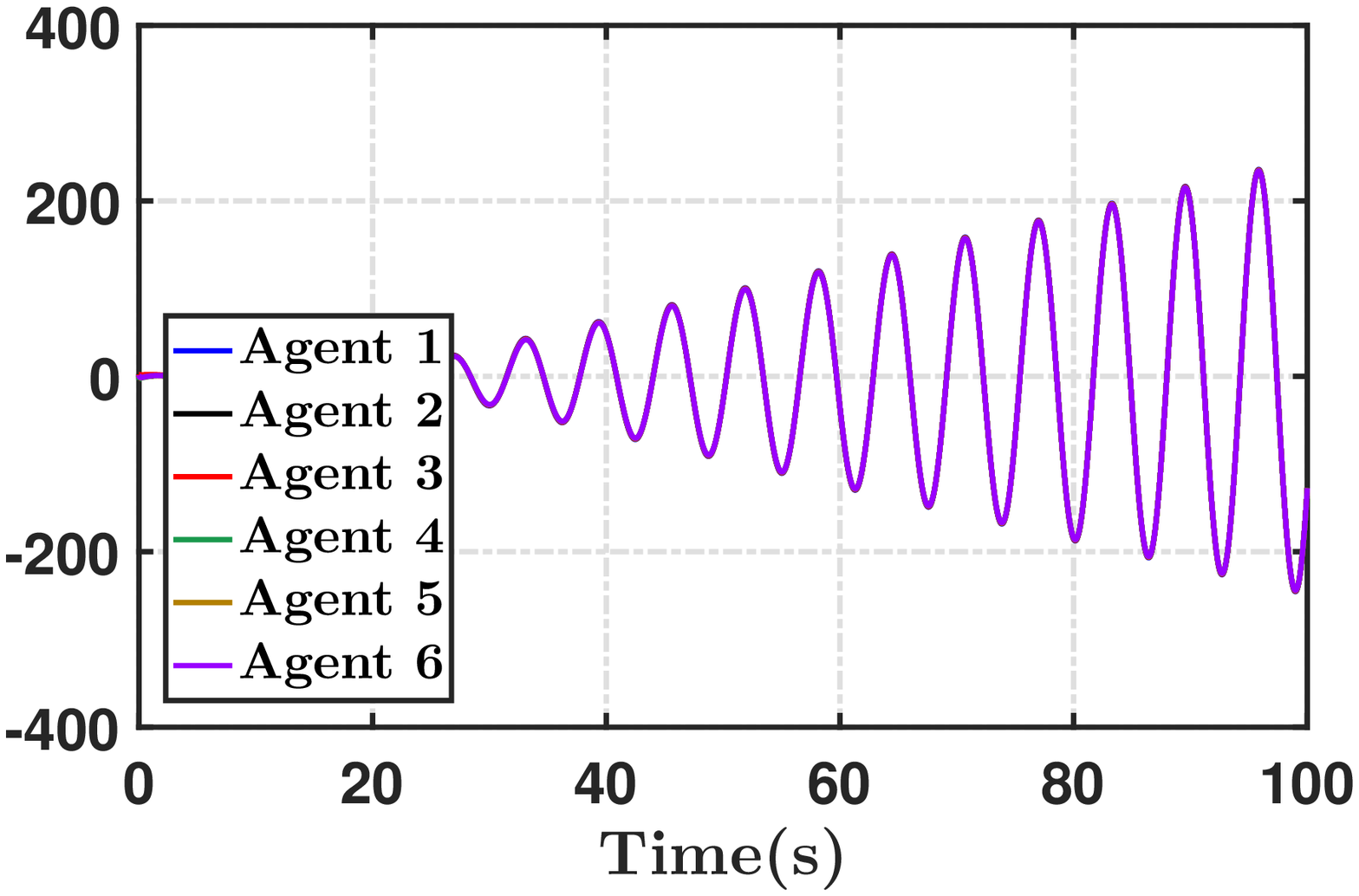}   
        \caption{}
        \label{fig:S_RS_UL}
    \end{subfigure}
    \hspace{0cm}
    \begin{subfigure}[b]{0.23\textwidth}
        \centering
        \includegraphics[width=1\linewidth,height=3.2cm]{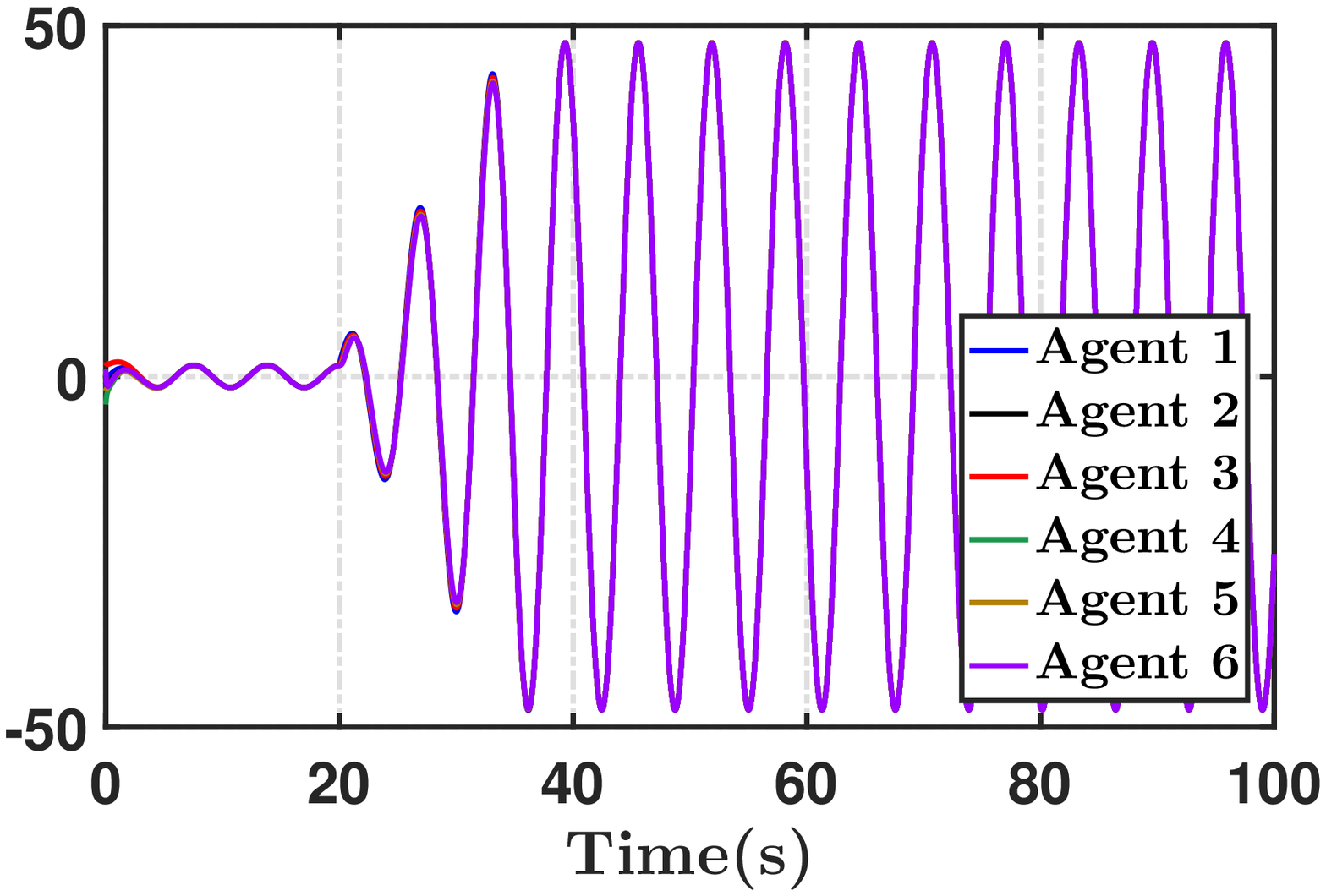}
        \caption{}
        \label{fig:S_RS_L}
    \end{subfigure}
    \begin{subfigure}[b]{0.23\textwidth}
        \centering
        \includegraphics[width=1\linewidth,height=3.2cm]{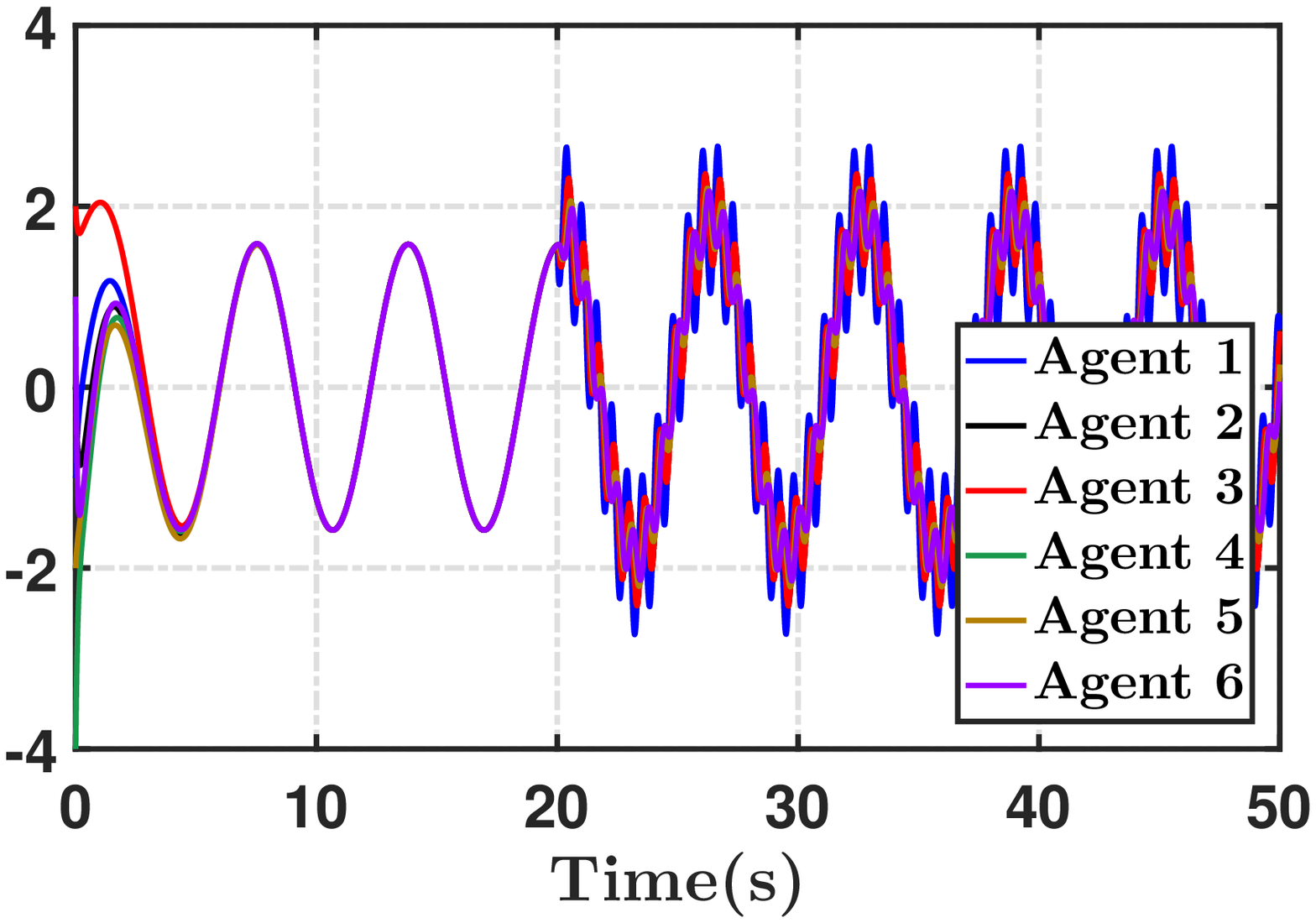}   
        \caption{}
        \label{fig:S_RD_UL}
    \end{subfigure}
    \hspace{0cm}
    \begin{subfigure}[b]{0.23\textwidth}
        \centering
        \includegraphics[width=1\linewidth,height=3.2cm]{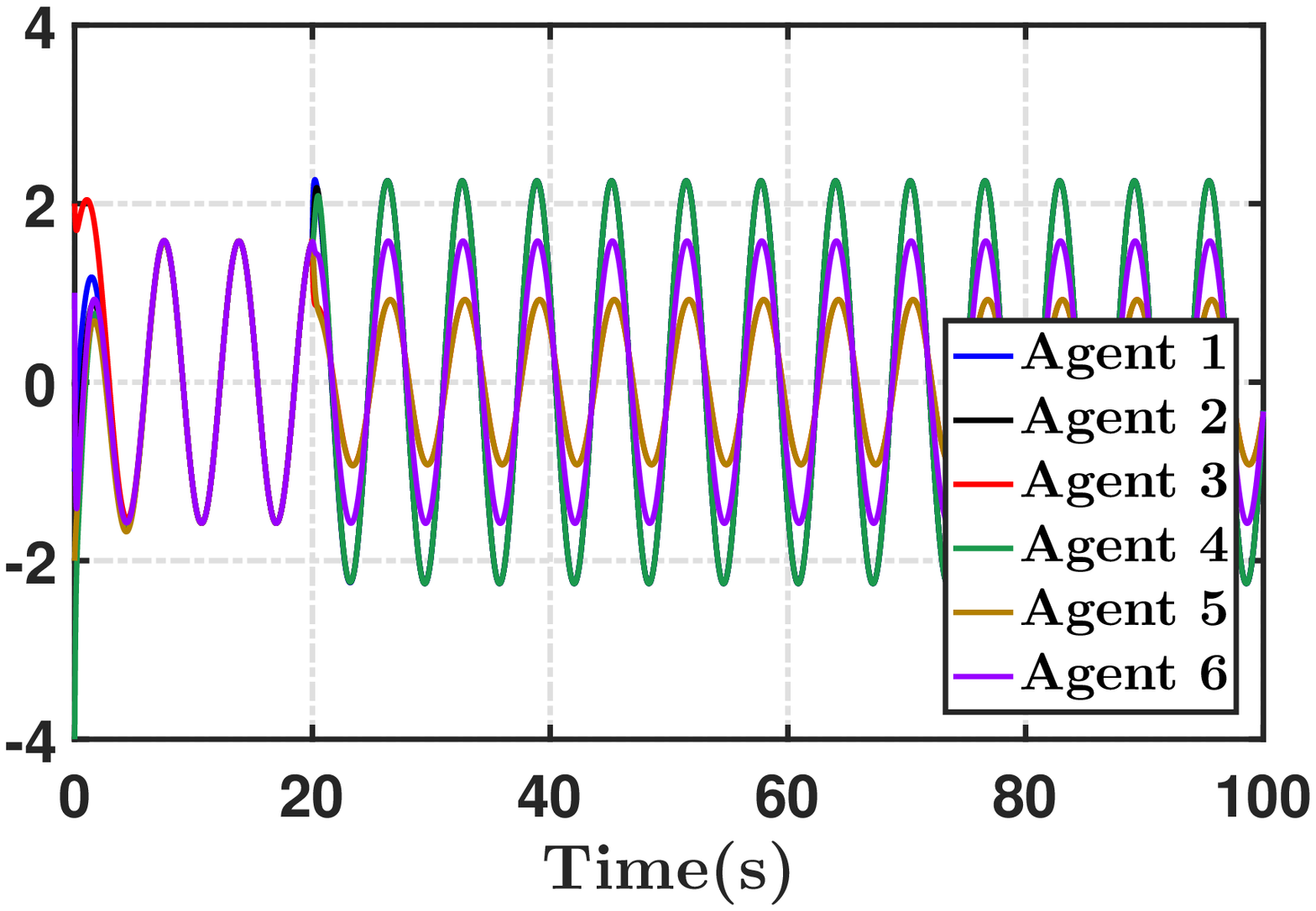}
        \caption{}
        \label{fig:S_RS_T3}
    \end{subfigure}
    \caption{Agent's state, when attack signal is injected into Agent 1 (Root node). (a) Agent 1 is under $f_1$. (b) Agent 1 is under $f_2$. (c) Agent 1 is under $f_3$. (d) Agent 1 and Agent 3 is under $f_1$ and $(-f_1)$.}
    \label{fig:RS_UL}
\end{figure}

Assume that attack signal defined in \eqref{eq:57} is injected into Agent 1. The agent's state are shown in Fig. \ref{fig:RS_UL}. Fig. \ref{fig:S_RS_UL} reveals that the entire network converge to infinity for the long-duration attack signal $f_1$. It can be seen from Fig. \ref{fig:S_RS_L} that for the short-duration attack $f_2$, once the attack is removed, all agents reach consensus but to a wrong trajectory because the attacker has changed the state of root nodes and, therefore, network shows a misbehavior. Fig. \ref{fig:S_RD_UL} shows that under the attack signal $f_3$ all agents in the network show a non-emergent behavior. However, agents do not converge to infinity. These results are comply with Theorems 1, 2 and 4. Now, let Agent 1 be under attack signal $f_1$ and Agent 3 is under attack signal $(-f_1)$ such that condition \eqref{eq:T3_eq1} in Theorem 3 holds. It can be seen from Fig. \ref{fig:S_RS_T3} that agents do not converge to infinity, however, because of the attack effect, the network shows a non-emergent behavior. This is consistent with Theorems 2 and 3.

\subsection{Attack on non-Root Node}
In this subsection, the effects of the attack signal on non-root nodes are presented.

\begin{figure}
    \centering
    \begin{subfigure}[b]{0.23\textwidth}
        \centering
        \includegraphics[width=1\linewidth,height=3.2cm]{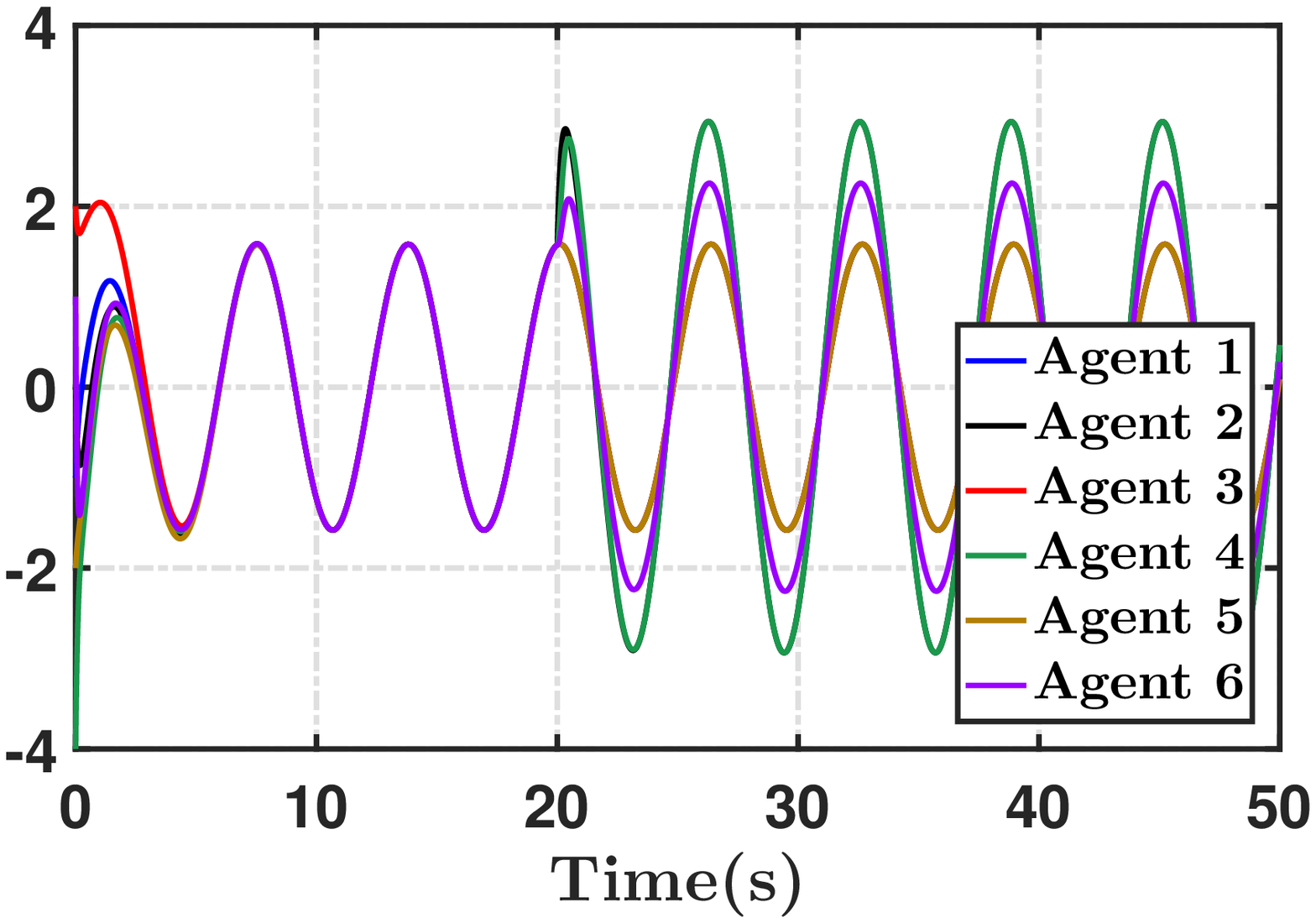}   
        \caption{}
        \label{fig:S_NRS_U}
    \end{subfigure}
    \hspace{0cm}
    \begin{subfigure}[b]{0.23\textwidth}
        \centering
        \includegraphics[width=1\linewidth,height=3.2cm]{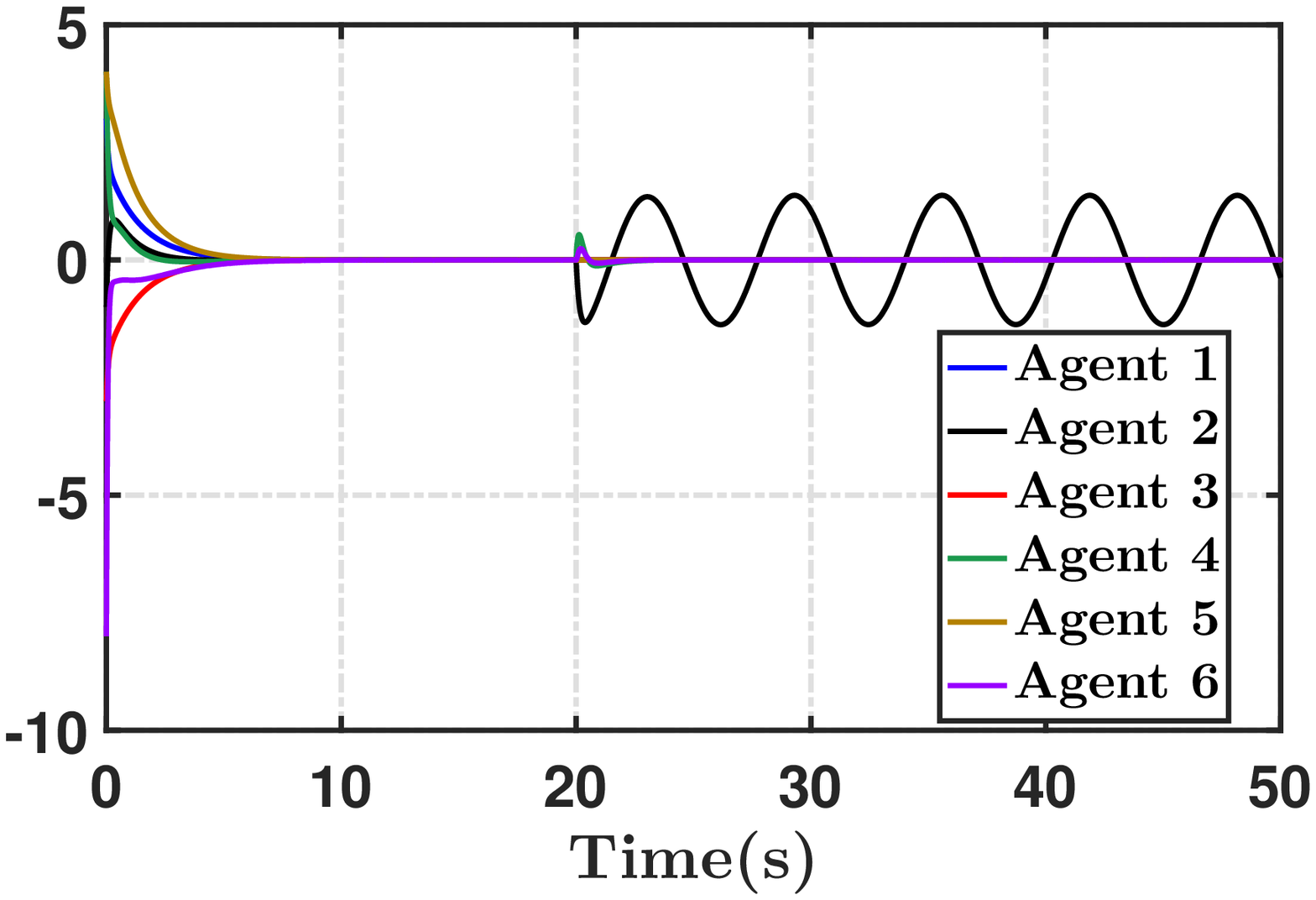}
        \caption{}
        \label{fig:E_NRS_U}
    \end{subfigure}
    \begin{subfigure}[b]{0.23\textwidth}
        \centering
        \includegraphics[width=1\linewidth,height=3.2cm]{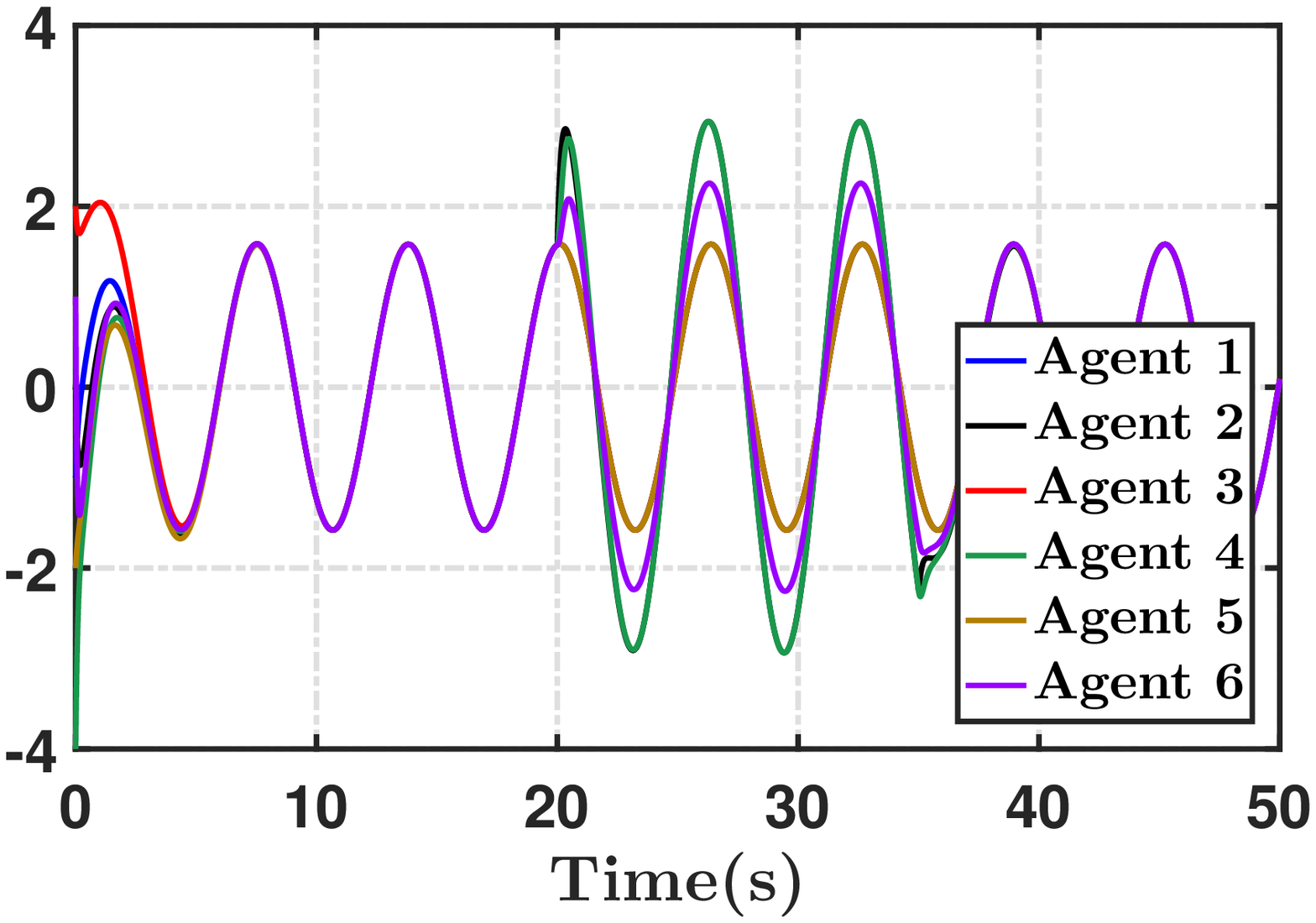}   
        \caption{}
        \label{fig:S_NRS_L}
    \end{subfigure}
    \hspace{0cm}
    \begin{subfigure}[b]{0.23\textwidth}
        \centering
        \includegraphics[width=1\linewidth,height=3.2cm]{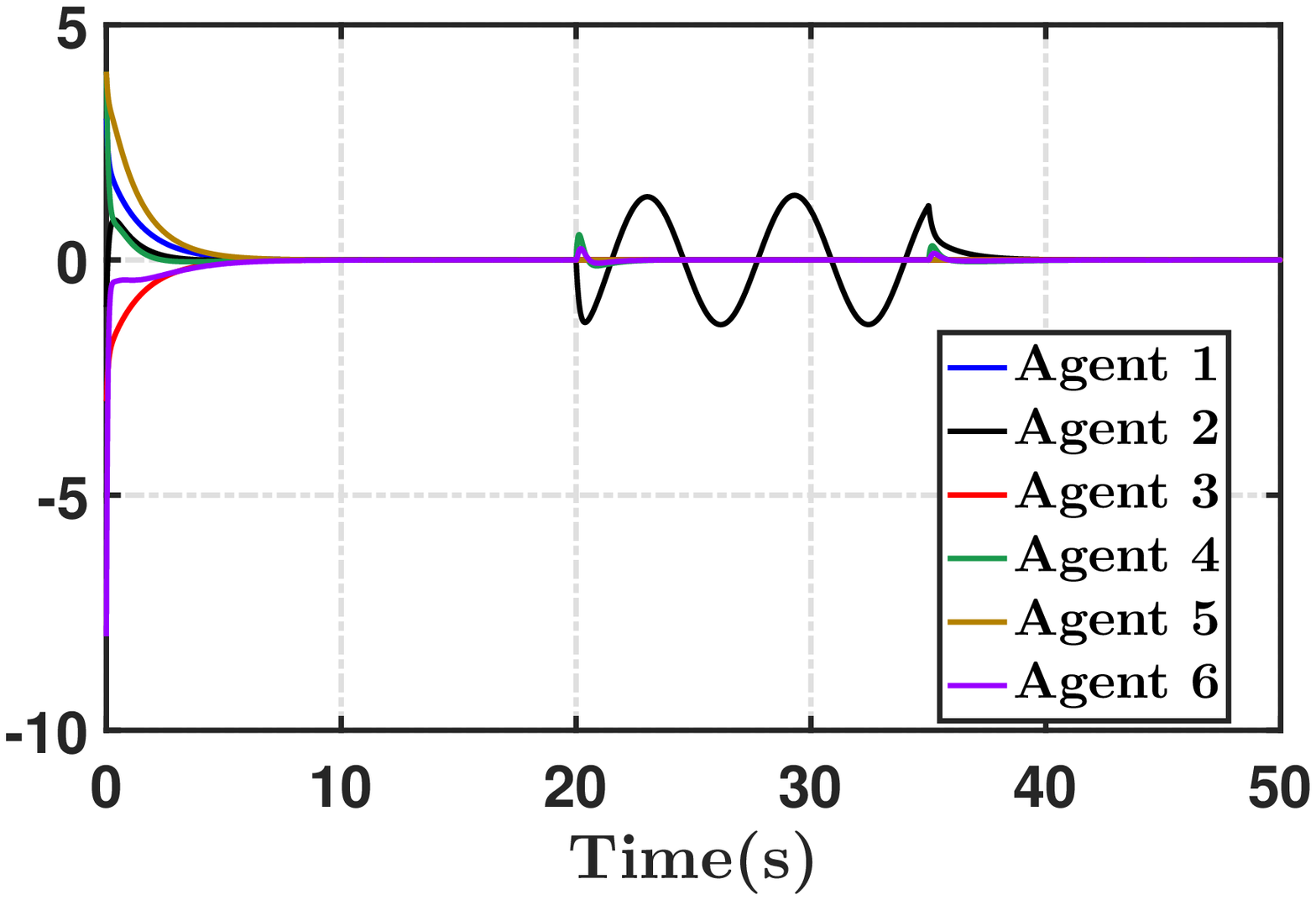}
        \caption{}
        \label{fig:E_NRS_L}
    \end{subfigure}
    \caption{The state and the local neighborhood tracking error of agents, when Agent 2 (non-Root node) is under the attack signal. (a) Agent 2 is under the attack signal $f_1$. (b) Agent's local neighborhood tracking error when the attack signal is $f_1$ (c) Agent 2 is under the attack signal $f_2$. (d) Agent's local neighborhood tracking error when the attack signal is $f_2$.}
    \label{fig:NRS_U}
\end{figure}

Let Agent 2 be under the attack signal defined in \eqref{eq:57}. The state and the local neighborhood tracking error of agents is shown in Fig. \ref{fig:NRS_U}. It is shown in Fig. \ref{fig:S_NRS_U} that Agents 2, 4 and 6 do not synchronize to the desired consensus trajectory. This is comply with Theorem 1. However, as shown in Fig. \ref{fig:E_NRS_U}, the local neighborhood tracking error for Agents 4 and 6 converges to zero and it is nonzero only for the compromised Agent 2. This is consistent with the result of Lemma 6 and Theorem 5. Now, let agent 2 be affected by the attack signal $f_2$. It can be seen in Fig. \ref{fig:S_NRS_L} and Fig. \ref{fig:E_NRS_L} that when the attack signal is removed, Agents 2, 4 and 6 synchronize to the desired trajectory, and the local neighborhood tracking error converges to zero. 

\section{Conclusion}
In this paper, we analyze the adverse effects of an attacker on the leaderless MAS. It is shown that a compromised agent propagates its adverse effects to all intact agents that are reachable from it. Furthermore, we show that the attacker can destabilize the whole synchronization process by injecting only a state-independent attack signal to sensors or actuators of a single root node or to its outgoing communication links. The attacker does not need to have any knowledge of the communication network or agents’ complete dynamics to destabilize the network. It only needs to  identify one of system eigenvalues by eavesdropping and monitoring the transmissions data to make the entire network unstable. We call this the internal model principle for the attacker. These attacks cannot be identified using existing model-based approaches and can destabilize the network quickly, and thus, practically halt the whole synchronization process. The importance of recognizing and characterizing these attacks is that one can empower the agents with resilient controllers to diminish them.

\bibliographystyle{plain}        

\end{document}